\documentclass[12pt]{JHEP3}

\usepackage{amssymb,amsmath}
\usepackage[mathscr]{eucal}
\usepackage{multicol}
\usepackage{amsfonts}
\usepackage{longtable}
\usepackage{rotating}

\usepackage{graphicx}
\newcommand{\be}{\begin{equation}}
\newcommand{\ee}{\end{equation}}
\newcommand{\bea}{\begin{eqnarray}}
\newcommand{\eea}{\end{eqnarray}}

\def\ket#1{|#1\rangle}

\newcommand*{\ar}[2]{\left(\frac{#1}{#2}\right)}
\def\cors#1{\langle\langle\, #1 \,\rangle\rangle}
\newcommand*{\R}[3]{\mathcal R^R_{#1}(#2|#3,1,a,b)}
\newcommand*{\Rpg}[3]{\mathcal R_{#1}(#2|#3,\lambda)}
\newcommand*{\Rpl}[3]{\mathcal R_{#1}(#2|#3,1)}

\def\fraction#1#2{ { \scriptstyle \frac{#1}{#2} }}

\def\p2{\fraction{\pi}{2}}

\title{
Pure Gauge Configurations and\\
Tachyon Solutions to String Field Theories Equations of Motion\\ }
\author{Irina~Ya.~Aref'eva, Roman~V.~Gorbachev \\
Steklov Mathematical Institute of RAS,
Gubkin st. 8, 119991, Moscow, Russia, \\
E-mail: \email{arefeva@mi.ras.ru}, \email{rgorbachev@mi.ras.ru}}
\author{Dmitry~A.~Grigoryev, Pavel~N.~Khromov, Maxim~V.~Maltsev \\
Faculty of Physics,
M.V.Lomonosov Moscow State University,
Leninskie Gory, 119991, Moscow, Russia, \\
E-mail: \email{dmitriy57@gmail.com}, \email{khromov.pn@gmail.com}, \email{mv.malcev@phys.msu.ru}}
\author{Peter~B.~Medvedev \\
Institute of Theoretical and Experimental Physics,
B. Cheremushkinskaya st. 25, 117218, Moscow, Russia, \\
E-mail: \email{pmedvedev@itep.ru}}

\date {\today}

\abstract{In construction of analytical solutions to open string
field theories pure gauge configurations parameterized by wedge
states play an essential role. These pure gauge configurations  are
constructed as perturbation expansions and to guaranty that these
configurations are asymptotical solutions to equations of motion
one needs to study convergence of the perturbation expansions.
We demonstrate that for the large parameter of the perturbation
expansion these pure gauge truncated configurations give divergent
contributions to the equation of motion on the subspace of the
wedge states. We perform this demonstration numerically for the pure
gauge configurations related to tachyon solutions for the bosonic and
NS fermionic SFT. By the numerical calculations we also show that the perturbation expansions are cured
by adding extra terms. These terms are nothing but the terms
necessary to make valued the Sen conjectures.}


\preprint{}

\begin{document}

\section{Introduction}

Finding nontrivial analytic solutions to classical string field theory (SFT)
is one of the long-standing  problems in string theory. The first nontrivial solution of the
SFT equation of motion has been found by Schnabl  \cite{Sch} for the
Witten open bosonic SFT  \cite{W}. The Schnabl paper has attracted a
lot of attention  \cite{Okawa}- \cite{Ishida:2008jc}.  It turns out that the
tachyon solution is closely related to pure gauge solutions. More
precisely, Schnabl's solution is a regularization of a singular
limit of a special pure gauge configuration  \cite{Sch,Okawa}. The
existence of pure gauge solutions to  the bosonic SFT equation of
motion  is provided by  the Chern-Simons form of the Witten cubic
action. The Schnabl solution is distinguished by its relation to a
true vacuum of the SFT, i.e. the vacuum on which the Sen
conjectures  \cite{Sen} are realized.

Schnabl's result has been generalized to the cubic super SFT
(SSFT)\footnote{We call the NS fermionic SFT the NS string field
theory with two sectors, $GSO(+)$ and $GSO(-)$ and we call the
superstring SFT (SSFT) the NS string field theory with the $GSO(+)$
sector only.}  \cite{AMZ,PTY} by Erler  \cite{Erler}. It is natural
to expect existence of a pure gauge solution to the cubic super SFT
(SSFT) equation of motion. However there is no reason for the
superstring case to deal with  the Sen conjecture, since the
perturbative vacuum is stable (there is no tachyon). However a
nontrivial (not pure gauge) solution to the SSFT equation of motion
does exist  \cite{Erler} \footnote{The physical meaning of this
solution is still unclear for us. It may happen that it is related
with a spontaneous supersymmetry breaking (compare with
\cite{AMZ2}).}.

To deal with the tachyon condensation  for the  fermionic string
one has to incorporate the $GSO(-)$ sector  \footnote{ Let us remind
that the NS fermionic SFT with two sectors is used to describe
non-BPS branes. The Sen conjecture has been checked by level
truncations for the non-polynomial and cubic cases in  \cite{BSZ} and
 \cite{ABKM}, respectively.}. A solution to the equation of motion of
the cubic SFT describing the NS string with both the $GSO(+)$ and
$GSO(-)$ sectors has been constructed in  \cite{AGM} (the AGM solution for short). For this
solution the first Sen conjecture has been checked analytically \cite{AGM}. A
solution to the equation of motion for the non-polynomial
SSFT  \cite{B} has been obtained in  \cite{Erler2,Okawa2}. This
construction became clear after a realization of an explicit
relation between the pure gauge solutions for the cubic superstring
field theories and non-polynomial ones found by Fuchs and Kroyter
 \cite{KF,KF-rev} (see also a recent discussion in  \cite{AGM-c}).

In  matrix notations  \cite{ABG} the equation of motion for the NS
fermionic SFT  has the form
 \be
 \label{EOM-m}
\widehat{Q}\widehat{\Phi}+\widehat{\Phi}\star\widehat{\Phi}=0, \ee
that is the same as the equation of motion for the open bosonic  and
cubic superstring SFTs (one has just removed the hats in the later
cases). $\widehat{\Phi}$ is $2\times2$ matrix where the components
are string fields belonging to the $GSO(+)$ and $GSO(-)$ sectors.
 Pure gauge solutions  can be written as
 \be
 \label{PG-m}
\widehat{\Phi}=-\widehat{Q}\widehat{\Omega}\star\widehat{\Omega}^{-1},
\ee where the parity of the entries of $\widehat{\Omega}$ must  be
adjusted  \cite{AGM}. For solution of Schnabl's form the following
formula is relevant \be \label{om} \widehat{\Omega}=\widehat 1-\lambda
\widehat{\varphi},\ee where $\widehat\varphi$ are special string
fields defining the choice of a solution.

In (\ref{PG-m})
$\widehat{\Omega}^{-1}$ is understood as a geometric series that
gives \be \widehat{\Phi}(\lambda)=\sum_{n=0}^\infty\lambda^{n+1}\widehat{Q}
\widehat{\varphi}\star\widehat{\varphi}^n.\ee For a special
$\widehat{\varphi}$ the pure gauge solution can be cast into the
form
\begin{equation}
\widehat\Phi(\lambda) =\sum_{n=0}^\infty\lambda^{n+1}\widehat\phi'_n \,,
\label{Psi-n}
\end{equation}
where the states $\widehat\phi'_n$ are defined for any real $n \ge 0$
and are made of the wedge state  \cite{Rastelli:2000iu,
Rastelli:2001vb, Schnabl2} (see explicit formula in section 2,
equations (\ref{matrix-comp}), (\ref{zeta0})-(\ref{eta}))

The Schnabl solution for the bosonic SFT as well as the Erler and
AGM solutions for the fermionic SFTs consist of two pieces. The
first piece is $\widehat{\Phi}(1)$. The second piece consists of the so
called phantom terms in the terminology of  \cite{KF}. The true solution to the equation of motion
is defined by the limit:
\begin{equation}
\widehat\Phi^R = \lim_{N \to \infty} \left[ \, \sum_{n=0}^N
\widehat\phi'_n - \widehat\psi_N \, \right]. \label{Psi}
\end{equation}

The phantom term for the Schnabl solution has appeared as an intrinsic part of the Schnabl
construction  \cite{Sch} \footnote{In  \cite{Okawa,FuchsKr} it has been
checked that the phantom term for the bosonic tachyon solution
provides the equation of motion contracted with the solution itself.}.
In the case of the Erler and AGM solutions phantom terms are added
 \cite{Erler,AGM} to satisfy the equation of motion contracted with
solutions themselves and to provide the Sen conjecture.

There is also another argument to add an extra term to the pure
gauge configurations. It is a convergency argument, or in other
words, a requirement that a perturbative solution must be also an
asymptotical weak solution on a subspace. As a subspace it is
reasonable to consider a subspace spanned by the wedge states. As it
was observed in  \cite{AGM-c} on the example of the cubic open SSFT,
 the perturbatively defined pure  gauge
configuration, related to the Erler solution,  fails to be a
solution to the equation of motion when contracted with wedge states. It has been shown that it is possible to cure  the
perturbation expansion by adding  extra phantom terms that are  just
the terms that have been used previously to provide that the
equation of motion contracted with the solution itself be satisfied
 \cite{Erler}. Similar numerical results for the bosonic string have
been reported  in  \cite{AGMM}.

The main purpose of this paper is to study the convergence of the
pure gauge configurations related to the tachyon solution
 \cite{AGM}  and to test the corresponding phantom terms.

 We show that
\begin{itemize}
\item
the pure gauge solution related to the AGM solution  and defined perturbatively,
\begin{equation}
\widehat\Phi_N(\lambda) = \sum_{n=0}^N \lambda^{n+1}\widehat\phi'_n \,,
\label{Psi-N}
\end{equation}
is divergent at $\lambda=1$ in the sense that the correlator

 \be
\label{EOM-Nm} \langle\langle\widehat\phi _m,\widehat Q\widehat \Phi _N
(1)+\widehat\Phi _N (1)\star\widehat\Phi _N(1)\rangle\rangle,
\ee
does not go to zero for any fixed $m$ and $N\to \infty$
and
\be
\label{EOM-m-lambda} \langle\langle\widehat\phi _m,\widehat Q\widehat \Phi
(\lambda)+\widehat\Phi (\lambda)\star\widehat\Phi (\lambda)\rangle\rangle \neq 0
 \,\,\,\mbox {for}
\,
|\lambda| \geq 1,
\ee
meanwhile
\be
\label{EOM-Nm} \langle\langle\widehat\phi _m,\widehat Q\widehat \Phi
(\lambda)+\widehat\Phi  (\lambda)\star\widehat\Phi (\lambda)\rangle\rangle =0
 \,\,\,\mbox {for}
\,
|\lambda|<1
\ee

\item
it is possible to cure the perturbation expansion
$\widehat\Phi_N(1)$ by adding extra terms $\widehat\psi_N$, \be
\widehat\Phi_N(1)\to
\widehat\Phi^R_N(1)\equiv\widehat\Phi_N(1)+\widehat\psi_N,\ee so
that the correlator
 \be
\label{EOM-Nm} \langle\langle\widehat\phi _m,\widehat Q\widehat \Phi ^R_N
(1)+ \widehat\Phi ^R_N (1)\star\widehat\Phi ^R_N(1)\rangle\rangle, \ee goes
to zero when $N\to \infty$.

\item $\widehat\psi_N$ is just the same term that has
been used previously to provide that the equation of motion
contracted with the solution itself be satisfied  \cite{AGM}.
\end{itemize}

The paper is organized as follows.

In Section 2 a matrix formulation for the NS fermionic SFT is
recalled and perturbative parameterizations of special pure gauge
configurations are presented. These pure gauge configurations are
used in the tachyon fermionic solution  \cite{AGM}.

Section 3 is devoted to the pure $GSO(+)$ sector and we give an
explicit demonstration  that $\lambda=1$ limit of the pure gauge
configurations used in the Erler construction  is in fact a singular
point and that it is possible to use a simple prescription to cure
divergences and this prescription gives the same answer as the
requirement of validity of the equation of motion contracted with
the solution itself.

Section 4 is devoted to the NS fermionic string including the $GSO(-)$
sector. In this case we do not have simple formulae for correlators
as we do have for the Erler solution. To demonstrate
 that $\lambda=1$ limit of the pure
gauge configurations  is in fact a singular point we use numerical
calculations. We also use numerical calculation to find phantom
terms that  cure  divergences. We show that the found phantom terms are
the same  as found before from
 the requirement of validity of the equation
of motion contracted with the solution itself.

In section 5 for completeness we collect the similar calculations
for the Schnabl solution to the bosonic SFT equation of motion
 adjusting the presentation of materials to our discussion  of the fermionic strings
 in section 3 and section 4.

In Appendix we collect correlators of wedge states with insertions used in the
construction of the solution to equation of motion.
\section{Set up}
\subsection{Perturbative Pure Gauge Solution}
\subsubsection{NS fermion string including $GSO(-)$ in matrix notations}
We begin with the  action  \cite{ABKM} (the ABKM action for short) in matrix
notations  \cite{ABG}
\begin{equation}\label{matrix-notations-action}
S[\widehat{\Phi}]=\frac12\langle\widehat{Y}_{-2}\widehat{\Phi},\widehat{Q}\widehat{\Phi}\rangle
+\frac13\langle
\widehat{Y}_{-2}\widehat{\Phi},\widehat{\Phi}\star\widehat{\Phi}\rangle,
\end{equation}
the string field $\widehat{\Phi}$ is given by
\begin{equation}\label{matrix-notations-field}
\widehat{\Phi}=\Phi_+\otimes\sigma_3+\Phi_-\otimes i\sigma_2,
\end{equation}
where $\Phi_+, \Phi_-$ are string fields which belong to $GSO(\pm)$  \cite{GSW} sectors respectively,
and
\begin{equation}\label{matrix-notations-BRST-and-Y}
\widehat{Q}=Q\otimes\sigma_3,\quad
\widehat{Y}_{-2}=Y_{-2}\otimes\sigma_3,
\end{equation}
$\sigma_i$ are Pauli matrices, $Q$ is the BRST charge and $Y_{-2}$ is a double step picture changing operator  \cite{ABGKM}.

The parity assignment and $\sigma_i$ algebra lead to the Leibnitz
rule
\begin{equation}\label{matrix-notations-Leibnitz}
\widehat{Q}(\widehat{\Phi}\star\widehat{\Psi})=
(\widehat{Q}\widehat{\Phi})\star\widehat{\Psi}+(-)^{|\widehat{\Phi}|}\widehat{\Phi}\star(\widehat{Q}\widehat{\Psi}),
\end{equation}
where
\begin{equation}
|\widehat{\Phi}|\equiv|\Phi_+|.
\end{equation}

The equation of motion in the matrix notations reads
\begin{equation}\label{matrix-notations-eom}
\widehat{Q}\widehat{\Phi}+\widehat{\Phi}\star\widehat{\Phi}=0.
\end{equation}
If $\widehat{\Phi}$ is a nontrivial solution for (\ref{matrix-notations-eom}),
then it
has to be Grassman  odd (i.e. $|\widehat{\Phi}|=1$). A pure gauge solution
to (\ref{matrix-notations-eom}) is
\begin{equation}
\label{matrix-notations-pure-gauge}
\widehat{\Phi}=-\widehat{Q}\widehat{\Omega}\star\widehat{\Omega}^{-1}.
\end{equation}
$\widehat{\Omega}$ is even,
(i.e. $|\widehat{\Omega}|=0$) and  in components has the form:
\begin{equation}
\widehat{\Omega}=\Omega_+\otimes I+\Omega_-\otimes\sigma_1.
\end{equation}
We parameterize $\widehat\Omega$  as  \cite{AGM}
\begin{equation}
\widehat{\Omega}=\widehat 1-\lambda\widehat\varphi,
\end{equation}
where
\begin{equation}\label{widephi}
\widehat{\varphi}=\varphi_+\otimes I+\varphi_-\otimes\sigma_1,
\end{equation}
$\varphi_+$ and $\varphi_-$ are components of the gauge field
$\widehat{\varphi}$ and they belong to the $GSO(+)$ and $GSO(-)$ sectors
respectively. The Grassman parities of $\varphi_+$ and $\varphi_-$ are
opposite.

For this $\widehat\Omega$ the pure gauge configuration
(\ref{matrix-notations-pure-gauge}) has the form
\begin{equation}\label{matrix-notations-PG-solution}
\widehat{\Phi}(\lambda)
=\lambda\widehat{Q}\widehat{\varphi}\star\frac{1}{1-\lambda\widehat{\varphi}}.
\end{equation}
One can expand the expression (\ref{matrix-notations-PG-solution})
in $\lambda$ to get
\begin{equation}\label{matrix-notations-PG-solution2}
\widehat{\Phi}(\lambda)
=\sum_{n=0}^\infty\lambda^{n+1}\widehat{Q}\widehat{\varphi}\star\widehat{\varphi}^n.
\end{equation}
In  \cite{AGM} has been proposed to take the following form for
$\varphi_+$ and $\varphi_-$
\begin{eqnarray}
\label{phi}
\varphi_+ &=&FBcF, \\
\label{psi}
\varphi_-&=&FB\gamma F.
 \end{eqnarray}
Here $\varphi_+$ and $\varphi_-$ are written in the split-string
notations. The detailed relation of the split-string formalism  and
CFT has been elaborated by Okawa and Erler  \cite{Okawa,Erler3} for
bosonic SFT and by Erler  \cite{Erler, Erler2} for SSFT. We sketch
the formulae related the split-string formalism and CFT in
Appendix.

The explicit expansion of the gauge configuration
(\ref{matrix-notations-PG-solution}) in the parameter $\lambda$ is

\begin{equation}
\widehat\Phi(\lambda) = \sum_{n=0}^\infty \lambda^{n+1}\widehat\phi'_n \,,
\label{Psi-n-m}
\end{equation}
where
\be
\label{matrix-comp}
\widehat{\phi}_n^\prime=\zeta_n^\prime\otimes\sigma_3+\xi_n^\prime\otimes
i\sigma_2.
\ee
In components
\be\label{matrix-lambda-field}
\widehat{\Phi}(\lambda)=\Phi_+(\lambda)\otimes\sigma_3+\Phi_-(\lambda)\otimes
i\sigma_2,
\ee
where
\begin{equation}
\label{component-notations-solution-GSO(+)}\Phi_+(\lambda)=\sum_{n=0}^\infty\lambda^{n+1}\zeta^\prime_{n},
\end{equation}
\begin{equation}
\Phi_-(\lambda)=\sum_{n=0}^\infty\lambda^{n+1}\xi^\prime_{n}.\label{component-notations-solution-GSO(-)}
\end{equation}

In the
split-string notations we have
 \cite{AGM}
\bea
\label{zeta0}\zeta'_0 &=& FcKBcF + FB\gamma^2F,\label{zeta0}\\
\xi^\prime_{0}&=&FcKB\gamma F +\frac12FB\gamma KcF +\frac12FB\gamma
cKF,\label{xi-0}\\
\label{zetan}\zeta'_n&=&\psi'_n+\chi'_n,\quad n>0,\\
\xi'_n&=&\vartheta'_n+\eta'_n,\,\,\quad n>0,
\eea
where
\begin{eqnarray}
\label{psi}\psi'_n&=&Fc\Omega^nKBcF,\quad n>0,\\
\label{chi}\chi'_n&=&F\gamma\Omega^nKB\gamma F,\quad n>0,\\
\label{vartheta}\vartheta'_n&=&F\gamma\Omega^nKBcF,\quad n>0,\\
\label{eta}\eta'_n&=&Fc\Omega^nKB\gamma F,\quad n>0.
\end{eqnarray}

\subsubsection{NS superstring}
To recover the Erler form  \cite{Erler} of the pure gauge
configuration $\Phi_S(\lambda)$ from
(\ref{matrix-notations-pure-gauge})-(\ref{component-notations-solution-GSO(+)})
one has to take $\varphi_-=0$ in (\ref{widephi}) and $\varphi_+$ as in
(\ref{phi}). This gives
\begin{equation}
\label{erl-pgc-exp}
\Phi_S(\lambda)=\sum_{n=0}^\infty\lambda^{n+1}\psi'_n,
\end{equation}
where
\begin{eqnarray}
\psi'_0&=&FcKBcF+FB\gamma^2F,\label{varphi0}\\
\psi'_n&=&Fc\Omega^nKBcF,\quad n>0 \label{varphi}.
\end{eqnarray}
Therefore, the first term in the R.H.S. of (\ref{zetan}) coincides with the
Erler term (\ref{varphi}) and the term (\ref{zeta0}) coincides with (\ref{varphi0}).

\subsubsection{Bosonic string}
Comparing  our results with the case of the bosonic SFT we use
the Schnabl form of the pure gauge configuration
\begin{equation}\label{Phi_B}
\Phi_B(\lambda)=\sum_{n=0}^\infty\lambda^{n+1}\varphi'_n,
\end{equation}
where
\begin{eqnarray}
\varphi'_0&=&FcKBcF,\\
\varphi'_n&=&Fc\Omega^nKBcF,\quad n>0.
\end{eqnarray}

\subsection{Asymptotic and Weak Asymptotic Solutions}

By construction the pure gauge configuration
(\ref{matrix-notations-PG-solution}) solves equation of motion
(\ref{matrix-notations-eom}) at each order in $\lambda$. However
this does not mean that the constructed pure gauge configuration has
a meaning within a non-perturbative framework. In particular, one
can wonder if the gauge configuration defines an asymptotic
solution to the equation of motion. Let us remind the definition of
asymptotic solution to the SFT equation of motion. $\widehat{\Phi}_N$ is called an
asymptotic solution to (\ref{matrix-notations-eom}), if
\begin{equation}
\lim_{N\to\infty}(\widehat Q\widehat{\Phi}_N+\widehat\Phi_N\star\widehat\Phi_N)=0.
\end{equation}

In our cases we can deal only with weak asymptotic solutions. Let us
remind $\widehat{\Phi}_N$ is called the weak asymptotic solution on a subspace
$\mathcal{S}$ if
\begin{equation}\label{asymp}
\lim_{N\to\infty}\langle\langle\psi,\widehat Q\widehat\Phi_N+\widehat\Phi_N\star\widehat\Phi_N\rangle\rangle=0
\end{equation}
for any $\psi\in\mathcal{S}$.

By construction we can guaranty that the pure gauge configuration is
a perturbative solution in a sense
\begin{equation}\label{eqm}
\widehat{Q}\widehat\phi'_n+\sum_{m=0}^n\widehat\phi'_m\star\widehat\phi'_{n-m}=0,
\end{equation}
but we cannot a priori guaranty that (\ref{asymp}) takes place.

\subsubsection{Notations for correlators}

We are going to consider the validity of equation of motion on a
subspace spanned by wedge state $\psi_n, \chi_n, \eta_n, \vartheta_n$
with $n>0$ and $\zeta'_0, \xi'_0$. We use the following
notations:
\begin{itemize}
\item for the NS fermion string case (the AGM pure gauge configuration)
\end{itemize}
\begin{align}
\label{RPG+} \Rpg{+}{\mathrm{field}}{N}&\equiv \cors{
\mathrm{field},Q\Phi_{+,N}(\lambda)+\Phi_{+,N}(\lambda)\star\Phi_{+,N}(\lambda)-\Phi_{-,N}
(\lambda)\star\Phi_{-,N}(\lambda)},\\
\Rpg{-}{\mathrm{field}}{N}&\equiv
\cors{\mathrm{field},Q\Phi_{-,N}(\lambda)+\Phi_{+,N}(\lambda)\star\Phi_{-,N}(\lambda)-\Phi_{-,N}(\lambda)\star\Phi_{+,N}(\lambda)},
\label{RPG-}
\end{align}
 where $\Phi_{\pm,N}(\lambda)$ are defined according to (\ref{component-notations-solution-GSO(+)}), (\ref{component-notations-solution-GSO(-)}) by
\begin{equation}
\begin{split}
\label{pg-agm-solution}
\Phi_{+,N}(\lambda)&=\sum_{n=0}^N\lambda^{n+1}\zeta^\prime_{n},\\
\Phi_{-,N}(\lambda)&=\sum_{n=0}^N\lambda^{n+1}\xi^\prime_{n};
\end{split}
\end{equation}

\begin{itemize}
\item for superstring case (the Erler pure gauge configuration)
\end{itemize}
\begin{equation}
\Rpg{S}{\mathrm{field}}{N}\equiv
\cors{ \mathrm{field},Q\Phi_{S,N}(\lambda)+\Phi_{S,N}(\lambda)\star\Phi_{S,N}(\lambda)},
\end{equation}
where $\Phi_{S,N}(\lambda)$ is given defined according to (\ref{erl-pgc-exp}) by
\begin{equation}
\Phi_{S,N}(\lambda)=\sum_{n=0}^N\lambda^{n+1}\psi'_n;
\end{equation}
\begin{itemize}
\item for the bosonic string (the Schnabl pure gauge configuration)
\end{itemize}
\begin{equation}
\label{pgc-b} \Rpg{B}{\mathrm{field}}{N}\equiv
\langle\mathrm{field},Q\Phi_{B,N}(\lambda)+\Phi_{B,N}(\lambda)\star\Phi_{B,N}(\lambda)\rangle,
\end{equation}
where $\Phi_{B,N}(\lambda)$ is given by
\begin{equation}\label{Phi_BN}
\Phi_{B,N}(\lambda)=\sum_{n=0}^N\lambda^{n+1}\varphi'_n.
\end{equation}
\section{Pure Gauge Configurations and the Erler Solution}
\subsection{Pure Gauge Configurations}

Let us check validity of the equation of motion  in weak sense on
the states $\psi_K$. For
this purpose we use the following correlators \footnote{From here
$\langle\langle...\rangle\rangle=\langle Y_{-2}...\rangle$.}
\begin{equation}
\begin{split}
\cors{\psi_K,Q\Phi_{S,N}} &= \frac{\lambda^2}{\pi^2}\frac{1-\lambda^N}{1-\lambda},\quad K>0,\\
\cors{\psi_K,\Phi_{S,N}\star\Phi_{S,N}} &= -\frac{\lambda^2}{\pi^2}\frac{1-\lambda^{N+1}}{1-\lambda},\quad K>0,\\
\cors{\psi'_0,Q\Phi_{S,N}} &= 0,\\
\cors{\psi'_0,\Phi_{S,N}\star\Phi_{S,N}} &= 0,\\
\end{split}
\end{equation}
to get
\begin{equation}
\begin{split}
\label{Erler-weak-EOM-pg}
\Rpg{S}{\psi_K}{N} &= -\frac{\lambda^{N+2}}{\pi^2},\quad K>0,\\
\Rpg{S}{\psi'_0}{N} &= 0.
\end{split}
\end{equation}

Taking the limit $N\to\infty$ for $\lambda<1$ we have for an
arbitrary $K>0$
\begin{equation}
\Rpg{S}{\psi_K}{\infty}=0.
\end{equation}
Note that for $\psi'_K$
\begin{equation}\label{der-psi-K}
\Rpg{S}{\psi'_K}{N}=\frac{d}{dK}\Rpg{S}{\psi_K}{N}=0,\quad K>0,
\end{equation}
in other words for $\lambda <1$  the field
$\Phi_{S,\infty}(\lambda)$ solves the equation of motion when
contracted with  states from the subspace spanned by $\psi_K$,
$\psi'_K$, $\psi'_0$. This fact is natural  for the solution
obtained by the iteration procedure (see section 4).

From equation \eqref{Erler-weak-EOM-pg} one sees  that for
$\lambda=1$ the string field  $\Phi_{S,N}(1)$ does not solve the
equation of motion even in weak sense
\begin{equation}
\label{SB-eom}
\mathcal R_S(\psi_K|N,1) = -\frac{1}{\pi^2}. 
\end{equation}

We have to note that using \eqref{Erler-weak-EOM-pg} and
\eqref{der-psi-K} we obtain that action on the pure gauge
configuration equals zero
\begin{equation}
S(\Psi_{S,N}(\lambda))\equiv 0.
\end{equation}

\subsection{The Erler Solution}
Let us add to $\Phi_{S,N}(1)\equiv \sum_{n=0}^{N}\psi_n'$ two extra terms
\begin{equation}
\label{reg2}
\Phi_{S,N}^R(a,b)=\Phi_{S,N}(1)+a\psi_N+b\psi_N',
\end{equation}
and find $a$ and $b$ from the requirement of validity of the equation of motion in weak sense
\begin{equation}
\begin{split}
\mathcal R_S^R(\psi_K|N,1,a,b)&=0,\quad K>0, \\
\mathcal R_S^R(\psi'_0|N,1,a,b)&=0.
\end{split}
\end{equation}
Here by the superscript $R$ we denote that we contract some field with equations of motion for
configuration with added phantom terms:
\begin{equation}
\mathcal R_S^R(\mathrm{field}|N,1,a,b)\equiv\cors{ \mathrm{field},Q\Phi_{S,N}^R+\Phi_{S,N}^R\star\Phi_{S,N}^R}.
\end{equation}

Simple calculations based on correlators \cite{Erler} show that
$a=-1$ and $b=-\frac12$. Indeed,
\begin{equation}
\begin{split}
\cors{\psi_K,Q\Phi_{S,N}^R(a,b)}&=\frac{(1+a)N+aK+2a+b}{\pi^2},\quad K>0,\\
\cors{\psi_K,\Phi_{S,N}^R(a,b)\star\Phi_{S,N}^R(a,b)}&=-\frac{(1+a)N+aK+3a+b+1}{\pi^2},\quad K>0,\\
\cors{\psi'_0,Q\Phi_{S,N}^R(a,b)}&=0,\\
\cors{\psi'_0,\Phi_{S,N}^R(a,b)\star\Phi_{S,N}^R(a,b)}&=-\frac{a(N+2+a(N+\frac32)+b)}{\pi^2}
\end{split}
\end{equation}
and we see that
\begin{equation}
\begin{split}
\R{S}{\psi_K}{N} &= -\frac{1+a}{\pi^2},\quad K>0, \\
\R{S}{\psi'_0}{N} &= -\frac{a(N+2+a(N+\frac32)+b)}{\pi^2}
\end{split}
\end{equation}
are equal to zero only for $a=-1$, $b=-\frac12$.

Next we calculate following correlators
\begin{eqnarray}
\cors{\Phi^R_{S,N},Q\Phi^R_{S,N}} &=& \frac{2a}{\pi^2}\Bigl[(1+a)N+a+b\Bigr],\\
\cors{\Phi^R_{S,N},\Phi^R_{S,N}\star\Phi^R_{S,N}} &=& -\frac{3a}{\pi^2}\left[(1+a)N+\frac 32 a+b+1\right].
\end{eqnarray}
Using these expressions we obtain contraction of the solution with the equation of motion
\begin{equation}
\cors{\Phi^R_{S,N},Q\Phi^R_{S,N}+\Phi^R_{S,N}\star\Phi^R_{S,N}} = -\frac{a}{\pi^2}\left[ (1+a)N +\frac 52 a+b+3 \right]
\end{equation}
equals zero when $a=-1$ and $b=-\frac 12$.
Following Erler  \cite{Erler} we can calculate action on this solution for arbitrary $a$ and $b$:
\begin{equation}
S = \frac 12\cors{\Phi^R_{S,N},Q\Phi^R_{S,N}}+\frac
13\cors{\Phi^R_{S,N},\Phi^R_{S,N}\star\Phi^R_{S,N}} =
-\frac{a(2+a)}{2\pi^2}.
\end{equation}
If we take $a=-1$ we get
\begin{equation}
S = \frac{1}{2\pi^2}.
\end{equation}
 It's interesting to note that $b$ does not
contribute to the value of action.

\section{Pure Gauge Configurations and the AGM Solution}
\subsection{Pure Gauge Configuration as Asymptotic Solution for Equation of Motion}
\begin{figure}[!t]
\begin{center}
\setlength{\unitlength}{1mm}
\begin{picture}(75,60)
\put(0,16){\begin{turn}{90}$\Rpg{+}{\zeta'_0}{N}$\end{turn}}
\put(5,-2){\includegraphics[height=6cm]{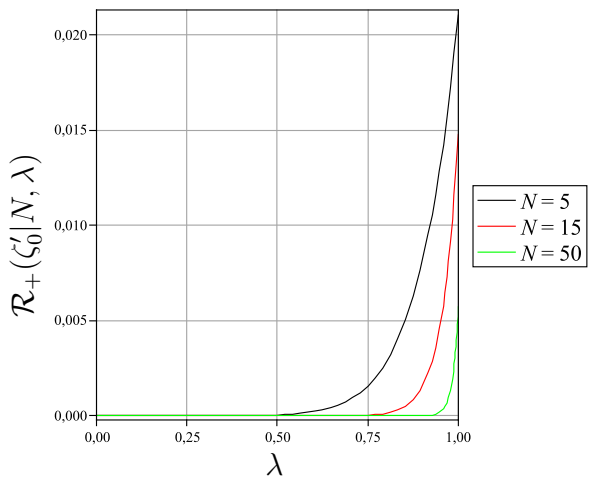}}
\put(34,-5){$\lambda$}
\end{picture}
\end{center}
\caption{Contraction
of equations of motion for $\Phi_{+,N}(\lambda)$, $N=5,10,15,20,50$ with $\zeta'_0$}
\label{R-phi0-N-lambda}
\end{figure}

Let us consider perturbative solution \eqref{pg-agm-solution} and
check whether it solves equations of motion in weak sense on the
subspace spanned by $\psi_n$, $\chi_n$, $\vartheta_n$, $\eta_n$ for
$n\geq1$ and $\zeta'_0$, $\xi'_0$ by direct computations. For this
purpose we evaluate expressions  (\ref{RPG+}) and (\ref{RPG-}) for
the mentioned fields. In figure  \ref{R-phi0-N-lambda} we see the
dependence of  $\Rpg{+}{\zeta'_0}{N}$ on $\lambda$ (recall that $N$
is the number at which we truncate the solution).  For the large $N$
the curve is closer to the $\lambda$-axis when $\lambda<1$. This
means that this perturbative  pure gauge configuration does in fact
solve the equation of motion in the weak sense.

\begin{figure}[!t]
\begin{center}
\setlength{\unitlength}{1mm}
\begin{picture}(70,60)
\put(0,16){\begin{turn}{90}$\Rpg{+}{\psi_1}{N}$\end{turn}}
\put(5,-2){\includegraphics[height=6cm]{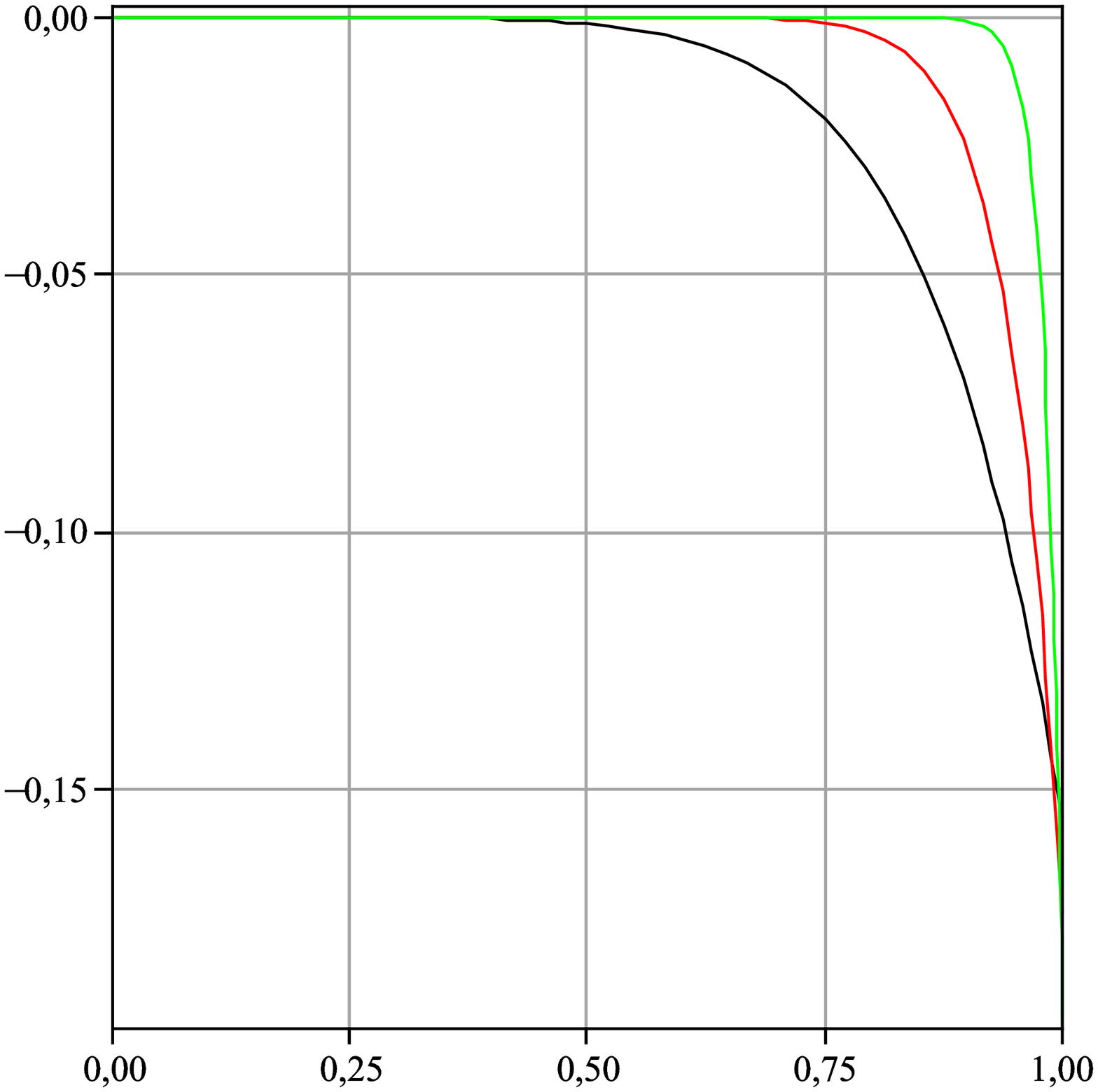}}
\put(36,-5){$\lambda$}
\end{picture}
\begin{picture}(75,60)
\put(0,16){\begin{turn}{90}$\Rpg{+}{\psi_{10}}{N}$\end{turn}}
\put(5,-2){\includegraphics[height=6cm]{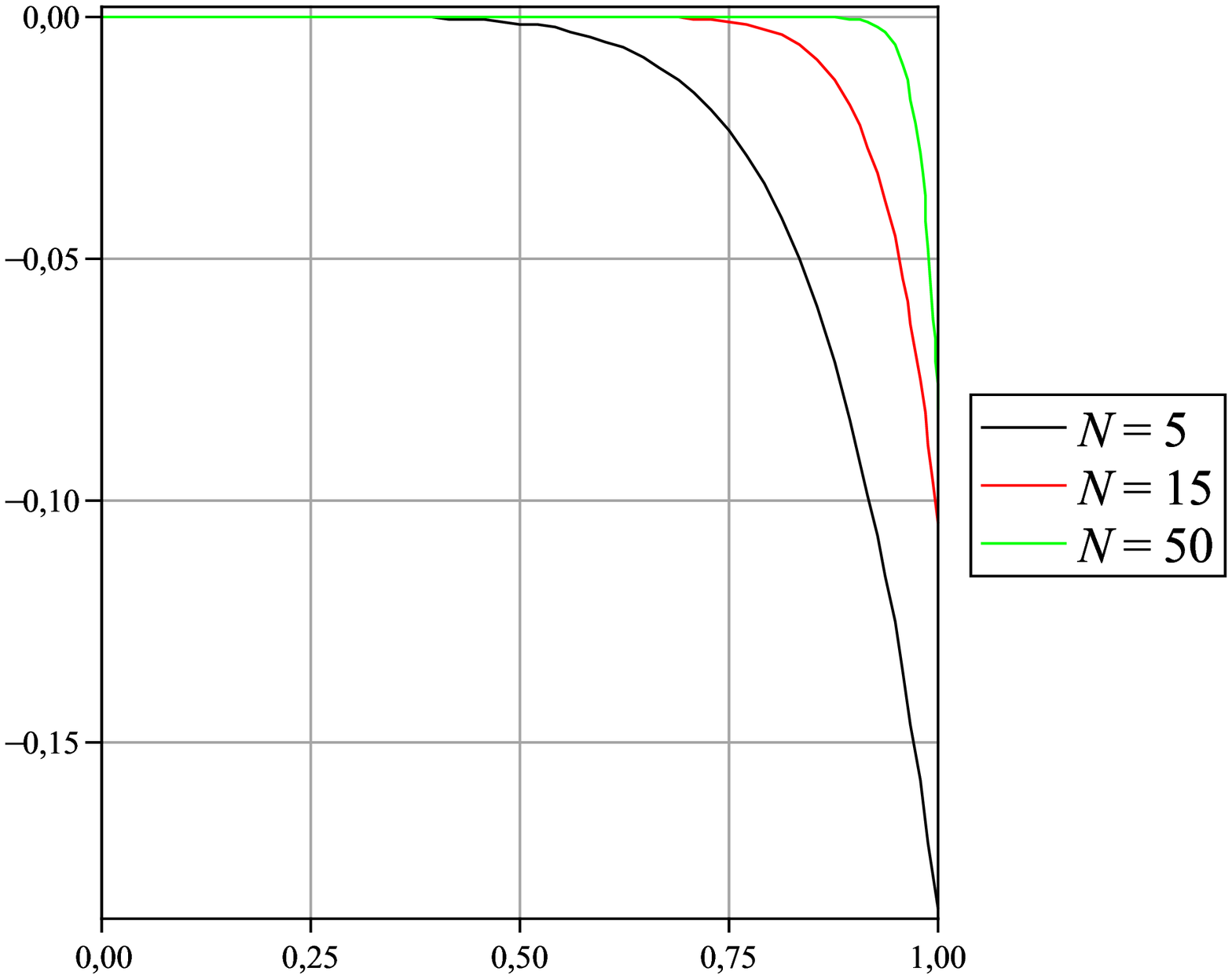}}
\put(36,-5){$\lambda$}
\end{picture}
\end{center}
\caption{Contraction of equations of motion for $\Phi_{+,N}(\lambda)$, $N=5,10,15,20,50$
with $\psi_1$ and $\psi_{10}$} \label{R-varphi-1-N-lambda}
\end{figure}

In figure  \ref{R-varphi-1-N-lambda} we see $\Rpg{+}{\psi_K}{N}$ for two
distinct values of $K$. We can see that for each $K$ the dependence on $\lambda$ and $N$
is similar and resembles $\Rpg{+}{\zeta'_0}{N}$. For contraction with other fields
($\chi_K$ and from $GSO(-)$ $\xi'_0$, $\vartheta_K$, $\eta_K$)
the results are similar, so we omit them.

Let us study more thoroughly equations of motion for pure gauge configuration at $\lambda=1$.
In figure  \ref{equation_without_ab} we can see that $\mathcal R_+(\psi_K|N,1)$ does not go to zero
with $N\to\infty$. We can make a claim that for different values of $K$ $\lim_{N\to\infty}\mathcal R_+(\psi_K|N,1)$
is the same. To support this claim we have calculated this expression at $N=1000$:
\begin{figure}[!t]
\begin{center}
\setlength{\unitlength}{1mm}
\begin{picture}(75,60)
\put(0,16){\begin{turn}{90}$\mathcal R_+(\psi_K|N,1)$\end{turn}}
\put(5,-2){\includegraphics[height=6cm]{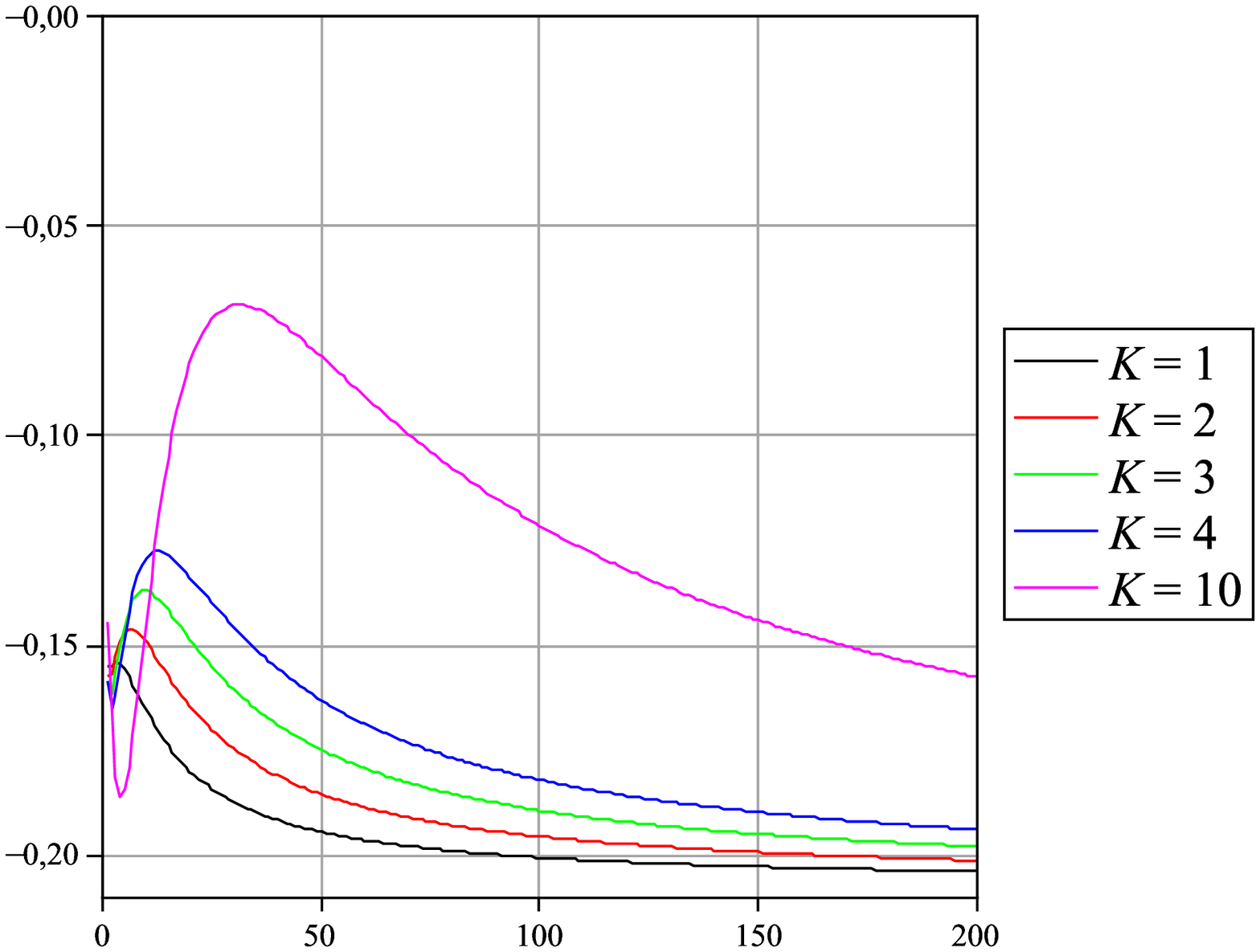}}
\put(34,-5){$N$}
\end{picture}
\end{center}
\caption{Contraction
of equations of motion for $\Phi_{+,N}(1)$ with $\psi_K$ for $K=1,2,3,4,10$}
\label{equation_without_ab}
\end{figure}

\begin{center}
\begin{tabular}{|c||c|c|c|}
\hline
$K$                             &       1   &        2  &       10 \\
\hline
$\mathcal R_+(\psi_K|1000,1)$   & -0.20649  & -0.20594  &  -0.19605 \\
\hline
\end{tabular}
\end{center}
Thus we claim that the limit is about $-0.2$ and the equations of motion do not
hold. This means that the pure gauge configuration $\Phi_{S,N}(\lambda)$ is
not an asymptotical solution at $\lambda=1$.

\subsection{Phantom Terms and the Equation of Motion on Some Low States}

We have seen that pure gauge configuration doesn't solve equations
of motion at $\lambda=1$, we need the phantom terms. In this
subsection we will show that both phantom terms with exactly the
coefficients written above are necessary to satisfy equations of
motion. First, we rewrite the partial solution with arbitrary
coefficients
\begin{equation}
\begin{split}
\label{AGM-solution-abcd}
\Phi_{+,N}^R(a,b) &= \sum_{n=0}^N\zeta'_n+a\zeta_N+b\zeta'_N,\\
\Phi_{-,N}^R(c,d) &= \sum_{n=0}^N\xi'_n+c\xi_N+d\xi'_N.
\end{split}
\end{equation}
To determine $a$ and $b$ coefficients it's enough to contract
equations of motion for $\Phi_{+,N}^R$ with $\zeta'_0$ and $\psi_1$
\begin{equation}\label{seek-for-ab}
\left\{
\begin{split}
\mathcal R_+^R(\zeta'_0|N,1,a,b)&=0, \\
\mathcal R_+^R(\psi_1|N,1,a,b)&=0.
\end{split}
\right.
\end{equation}
Here
\begin{equation}
\mathcal R_+^R(\mathrm{field}|N,1,a,b)=\cors{\mathrm{field},Q\Phi_{+,N}^R+\Phi_{+,N}^R\star \Phi_{+,N}^R-\Phi_{-,N}^R\star \Phi_{-,N}^R}.
\end{equation}

Similarly, $c$ and $d$ coefficients can be obtained by contracting
equations of motion for $\Phi_{-,N}$ with $\xi'_0$ and $\vartheta_1$
\begin{equation}\label{seek-for-cd}
\left\{
\begin{split}
\mathcal R_-^R(\xi'_0|N,1,c,d)&=0, \\
\mathcal R_-^R(\vartheta_1|N,1,c,d)&=0.
\end{split}
\right.
\end{equation}
Here
\begin{equation}
\mathcal R_-^R(\mathrm{field}|N,1,a,b)=\cors{\mathrm{field},Q\Phi_{-,N}^R+\Phi_{+,N}^R\star \Phi_{-,N}^R-\Phi_{-,N}^R\star \Phi_{+,N}^R}.
\end{equation}

By solving these equations numerically for several values of $N$ we obtain
the following results:
\begin{center}
\begin{tabular}{|c||c|c|c|c|}
\hline
$N$ &       $a$     &        $b$     &       $c$      &       $d$      \\
\hline \hline
1   & -1.009276758  & -0.4947687297  &  -1.021127911  &  -0.4547206603 \\
\hline
5   & -1.002887587  & -0.4802195028  &  -0.9958034182 &  -0.5258588923 \\
\hline
10  & -1.001361647  & -0.4839491795  &  -0.9996956796 &  -0.5033484993 \\
\hline
50  & -0.999973303  & -0.5013822672  &  -0.9999993154 &  -0.5000348047 \\
\hline
100 & -0.999996393  & -0.5003673241  &  -0.9999999396 &  -0.5000061487 \\
\hline
\end{tabular}
\end{center}
From this table we see that these values agree with exact values
$a=c=-1$ and $b=d=-\frac12$ with good precision.

\subsection{Phantom Terms and Equations of Motion on Higher States}
As we have seen in the previous subsection at $\lambda=1$ to have a solution of equations of motion in weak sense
on the four states \eqref{seek-for-ab}, \eqref{seek-for-cd} we have to add to the pure gauge solution two phantom terms
\begin{equation}
\begin{split}
\label{AGM-solution}
\Phi_{+,N}^R &= \sum_{n=0}^N\zeta'_n-\zeta_N-\frac 12\zeta'_N, \\
\Phi_{-,N}^R &= \sum_{n=0}^N\xi'_n-\xi_N-\frac 12\xi'_N.
\end{split}
\end{equation}
In this subsection we check that these terms also provide the
validity of equations of motion on higher states. For this purpose
we consider the following correlators
\begin{equation}
\mathcal R_+^R(\mathrm{field}|N,1,-1,-\frac12),\quad
R_-^R(\mathrm{field}|N,1,-1,-\frac12),
\end{equation}
where field is one of $\zeta'_0$, $\psi_K$, $\chi_K$, $\xi'_0$, $\vartheta_K$, $\eta_K$

\begin{figure}[!t]
\begin{center}
\setlength{\unitlength}{1mm}
\begin{picture}(70,60)
\put(0,16){\begin{turn}{90}$\mathcal R_+^R(\zeta'_0|N,1,-1,-\frac12)$\end{turn}}
\put(5,-2){\includegraphics[height=6cm]{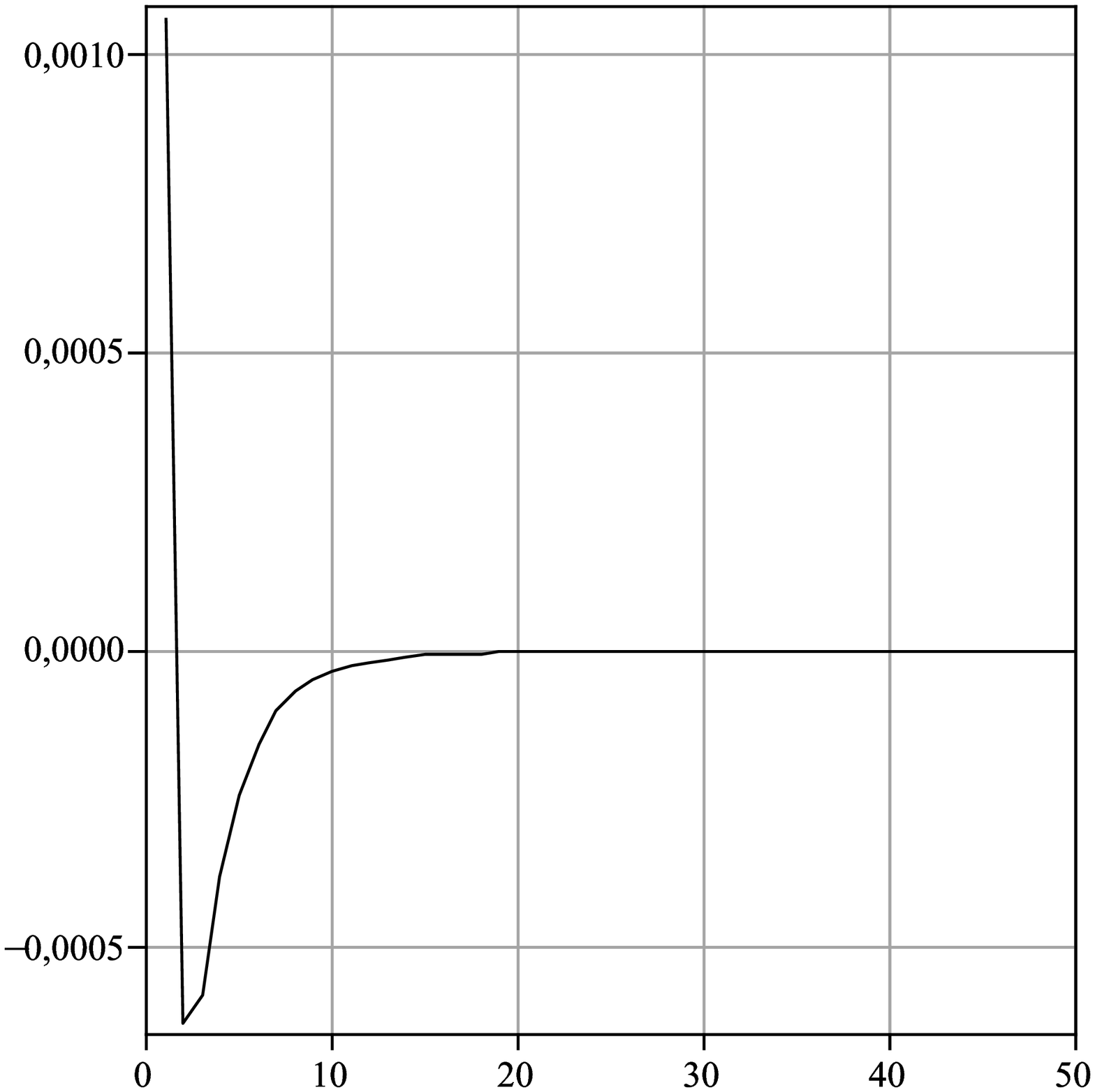}}
\put(36,-5){$N$}
\end{picture}
\begin{picture}(75,60)
\put(0,16){\begin{turn}{90}$\mathcal R_+^R(\psi_K|N,1,-1,-\frac12)$\end{turn}}
\put(5,-2){\includegraphics[height=6cm]{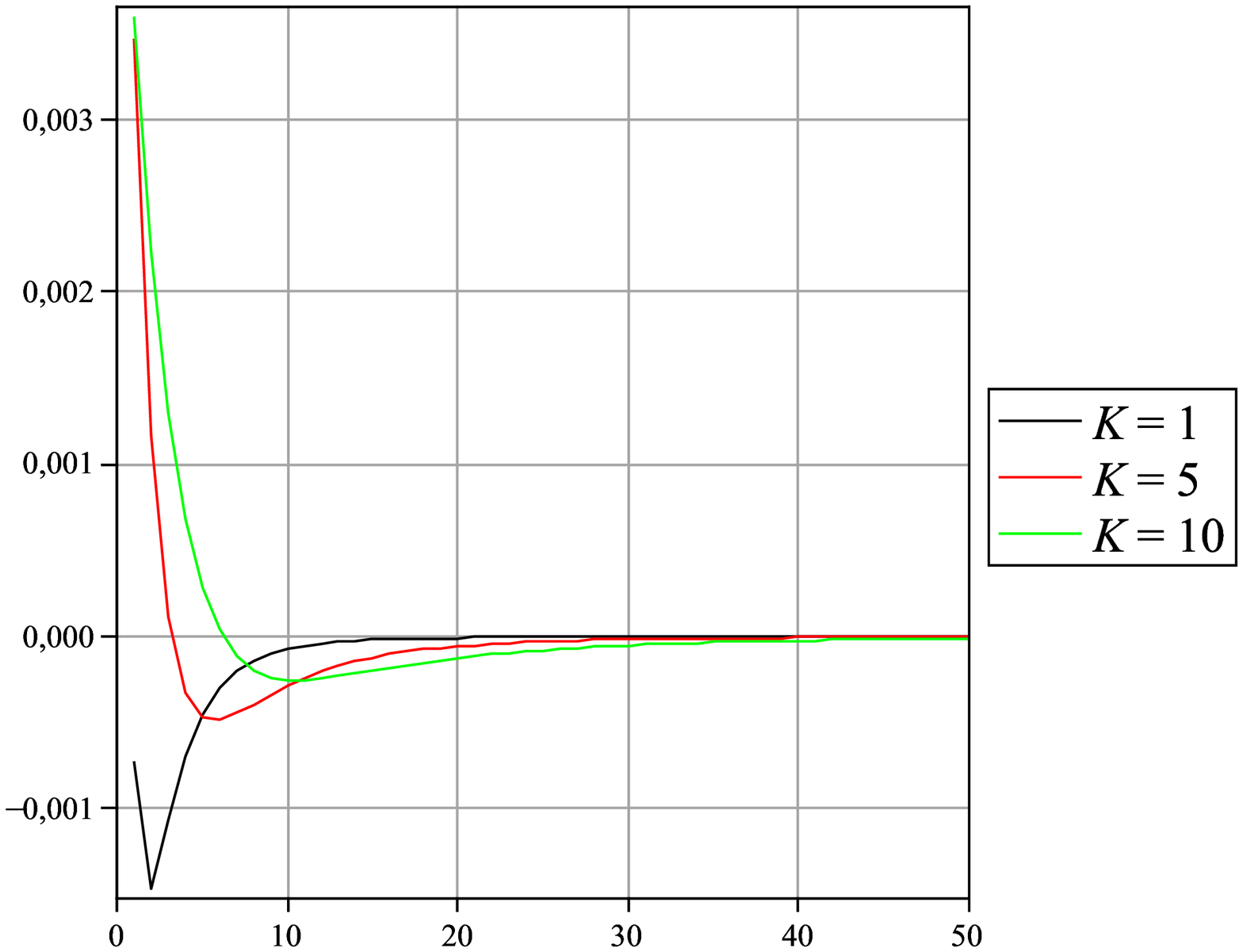}}
\put(36,-5){$N$}
\end{picture}
\end{center}
\caption{Contraction of equations of motion for $\Phi^R_{+,N}$ with $\zeta'_0$ and $\psi_K$ for
$K=1,5,10$}
\label{R-phi0-N-full}
\end{figure}

In figure  \ref{R-phi0-N-full} we see $\mathcal
R_+^R(\zeta'_0|N,1,-1,-\frac12)$ and $\mathcal
R_+^R(\psi_K|N,1,-1,-\frac12)$ for three values of $K$. We can see
that for large $N$ $\Phi^R_{+,N}$ asymptotically solves the equation
of motion. We also see that $\mathcal R_+^R(\psi_K|N,1,-1,-\frac12)$
as a function of $N$ has an extremum near $N=K$. We can think that
the largest contribution comes from correlators where $N\approx K$.
This means that $\mathcal R_+^R(\psi_K|N,1,-1,-\frac12)$ must be
small for $N\gg K$ because $\Phi_{+,\infty}^R$ is a solution. For
other fields we obtained the same results.

\subsection{Calculation of Action}

Let us calculate the action on the AGM solution. We calculate the
action on partial sums and present the result depending on index $N$
of partial sum. The action is expected to be $\frac{1}{2\pi^2}$, so
we multiply it by $2\pi^2$ to compare with $1$.

\begin{center}
\begin{tabular}{|c|c|}
\hline
$N$ & $2\pi^2S(\Phi_{+,N},\Phi_{-,N})$ \\
\hline
1   & 0.998538383 \\
\hline
5   & 0.999990263 \\
\hline
10  & 0.999999488 \\
\hline
15  & 0.999999919 \\
\hline
20  & 0.999999979 \\
\hline
30  & 0.999999997 \\
\hline
\end{tabular}
\end{center}

\begin{figure}[!t]
\begin{center}
\setlength{\unitlength}{1mm}
\begin{picture}(70,60)
\put(0,23){\begin{turn}{90}$2\pi^2S$\end{turn}}
\put(5,-2){\includegraphics[height=6cm]{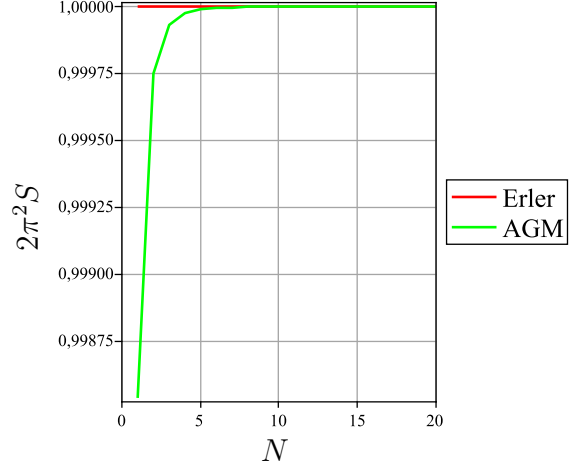}}
\put(33,-5){$N$}
\end{picture}
\end{center}
\caption{Dependence of action on $N$.}
\label{graph-Action-N}
\end{figure}

From the table above and from  figure  \ref{graph-Action-N} we see
that action has a good convergence to the expected value. This
proves the first Sen conjecture.

For comparison on the same figure we have presented the action on
the Erler solution. We can see that it doesn't depend on $N$. We
also see that both curves join asymptotically. This figure forces us
to conclude that $GSO(-)$ does not contribute to the value of action
and it provides a numerical confirmation of the claim given in
\cite{AGM}.

\section{Pure Gauge Configurations and the Schnabl Solution}
\subsection{Contractions for Pure Gauge Configurations}

\begin{figure}[!h]
\begin {center}
\setlength{\unitlength}{1mm}
\begin{picture}(130,60)
\put(0,0){\includegraphics[width=6cm]{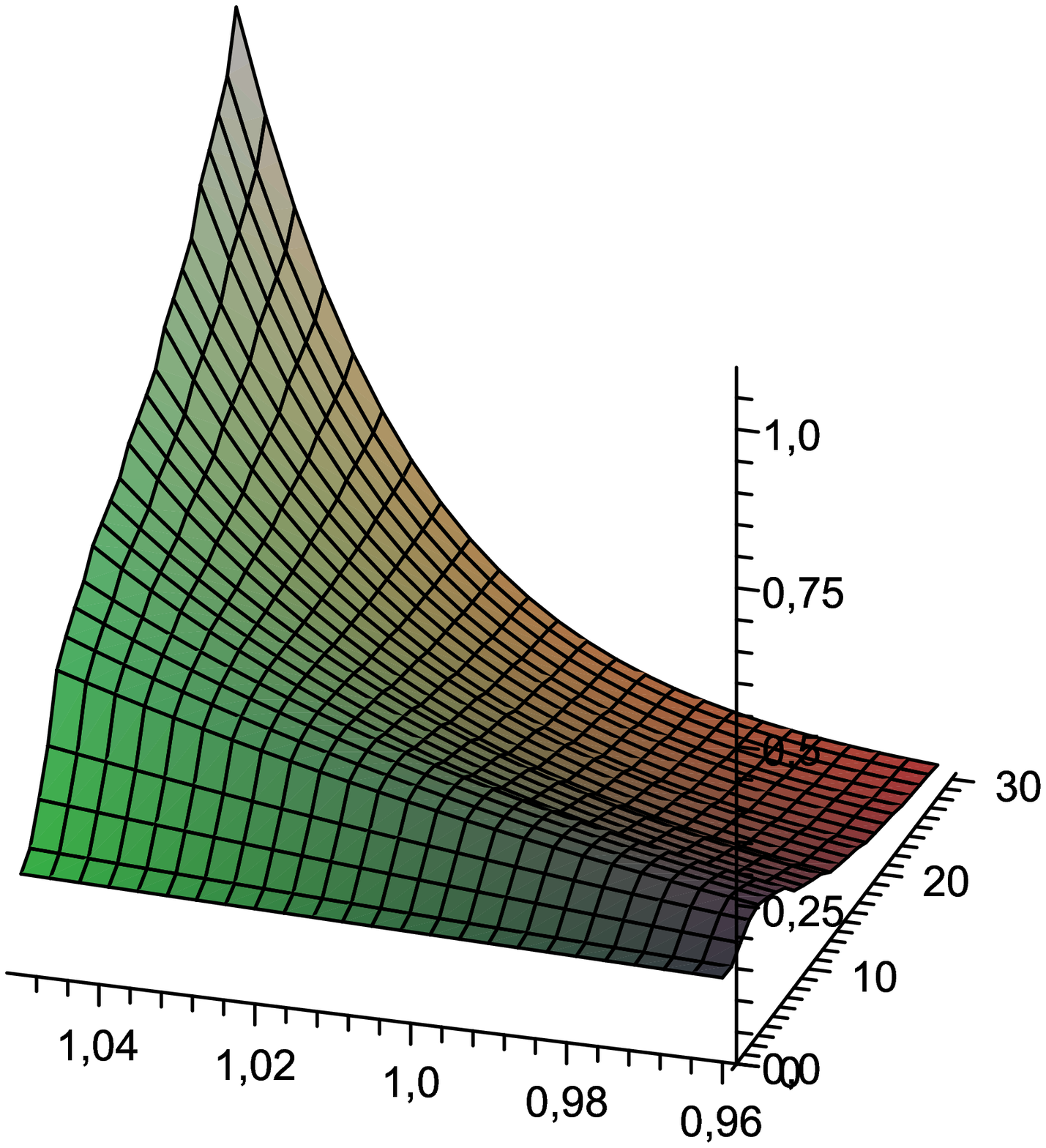}A.}
\put(18,3){\makebox(0,0)[lb]{$\lambda$}}
\put(52,10){\makebox(0,0)[lb]{$N$}}
\put(30,40){\makebox(0,0)[lb]{$\mathcal R_B(\varphi_1 |
N,\lambda)$}}
\put(65,0){\includegraphics[width=6cm]{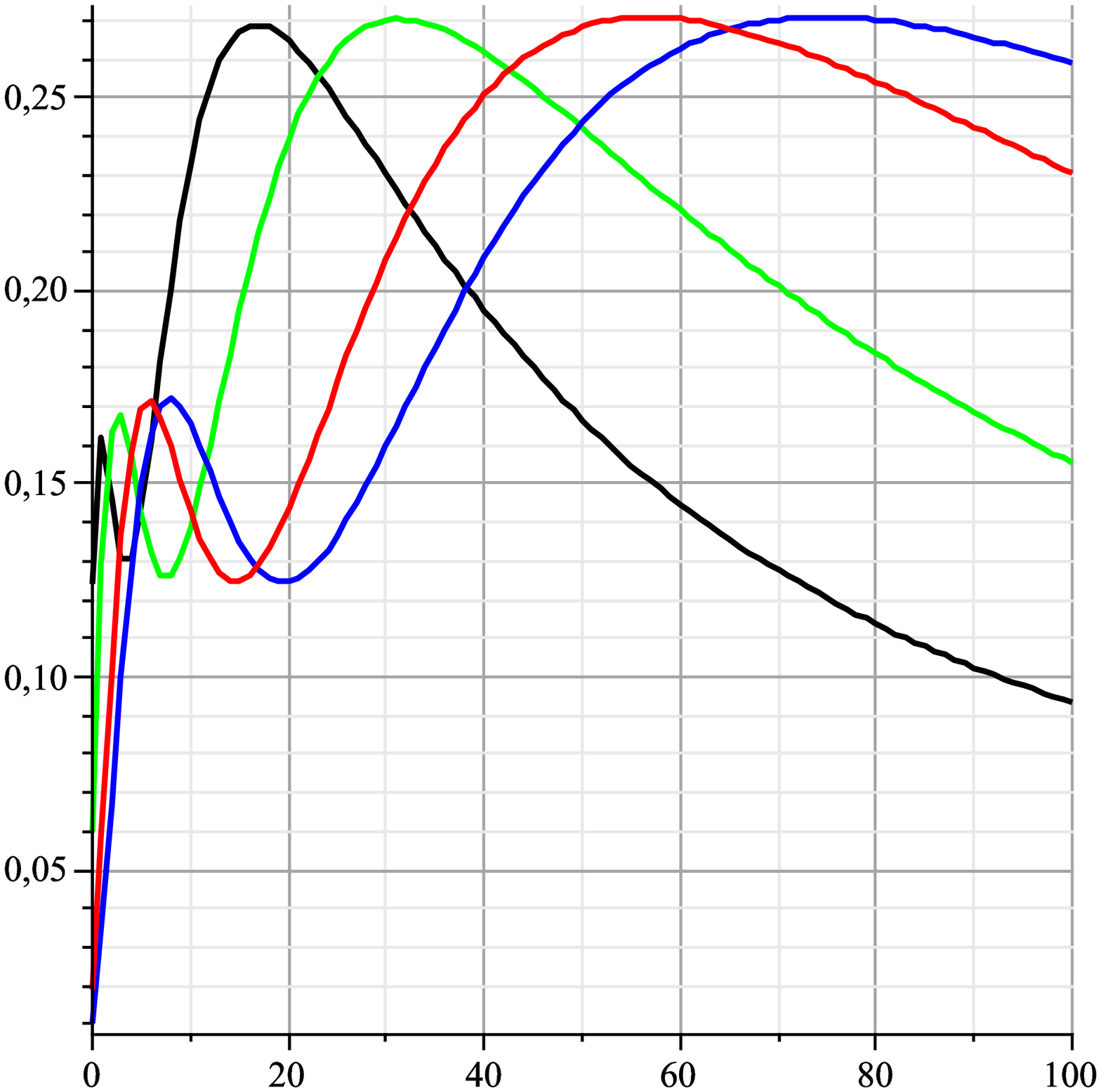}B.}\label{fig-kkk}
\put(95,-3){\makebox(0,0)[lb]{$N$}} \put(60,17){\begin{sideways}
{$\mathcal R_B(\varphi_K | N,1)$} \end{sideways}}
\put(123,20){\makebox(0,0)[lb]{\small{$K=$4}}}
\put(123,33){\makebox(0,0)[lb]{\small{$K=$8}}}
\put(123,49){\makebox(0,0)[lb]{\small{$K=$15}}}
\put(123,55){\makebox(0,0)[lb]{\small{$K=20$}}}
\end{picture}
\end{center}
\caption{\newline A. Contraction of equation of motion for
$\Phi_{B,N}(\lambda)$ with $\varphi_1$, $ N\leq 30$,
$0.96<\lambda<1.06$,\newline B. Contraction of the equation of motion for
$\Phi_{B,N}(\lambda)$ with $\varphi_{K}$, $1\leq N\leq 100$,
$\lambda=1$ and $K=4,8,15,20$} \label{Rb1}
\end{figure}

Let us for completeness consider the Schnabl pure gauge
configuration \eqref{Phi_B} that has been already done in many
details \cite{Sch},  \cite{Okawa},  \cite{FuchsKr}. To check by
direct computations whether \eqref{Phi_BN} solves equations of
motion in weak sense on the subspace spanned by $\varphi_K$  we
evaluate expression (\ref{pgc-b}). The results of calculations are
presented in figure \ref{Rb1}. Here $\Rpg{B}{\varphi_1}{N}$  is
shown as a function of $N$ and $\lambda$.
 We  see that for the large  $N$ and $\lambda< 1$
the surface is closer to the $(\lambda,N)$-plane meanwhile  when
$\lambda>1$ the surface blow up. A similar picture one gets for any
$K$ (see   \cite{AGMM} for more details). For $\lambda=1$ we observe
numerically an interesting phenomena that for any $N$ there is a
specific $K=K_{max}(N)$ so that
$|\Rpl{B}{\varphi_{K_{max}(N)}}{N}|>R_0$, where $R_0$ is an
universal constant and one can see that $R_0>0.26$ (see figure 6.B).

This means that the perturbative pure gauge configuration does not
solve the equation of motion in the  uniformly weak sense on a
subspace spanned by $\varphi _K$ for $\lambda=1$ but does solve it
in the weak sense. A similar picture we get for contractions with
$\varphi _K'$. This fact explains the observation
\cite{Okawa,FuchsKr} that
  the perturbative pure gauge configuration does not solve
the equation of motion contracted with this configuration itself. We
can also demonstrate this numerically.

\begin{figure}[!b]
\begin {center}
\setlength{\unitlength}{1mm}
\begin{picture}(130,60)
\put(0,0){\includegraphics[width=5 cm]{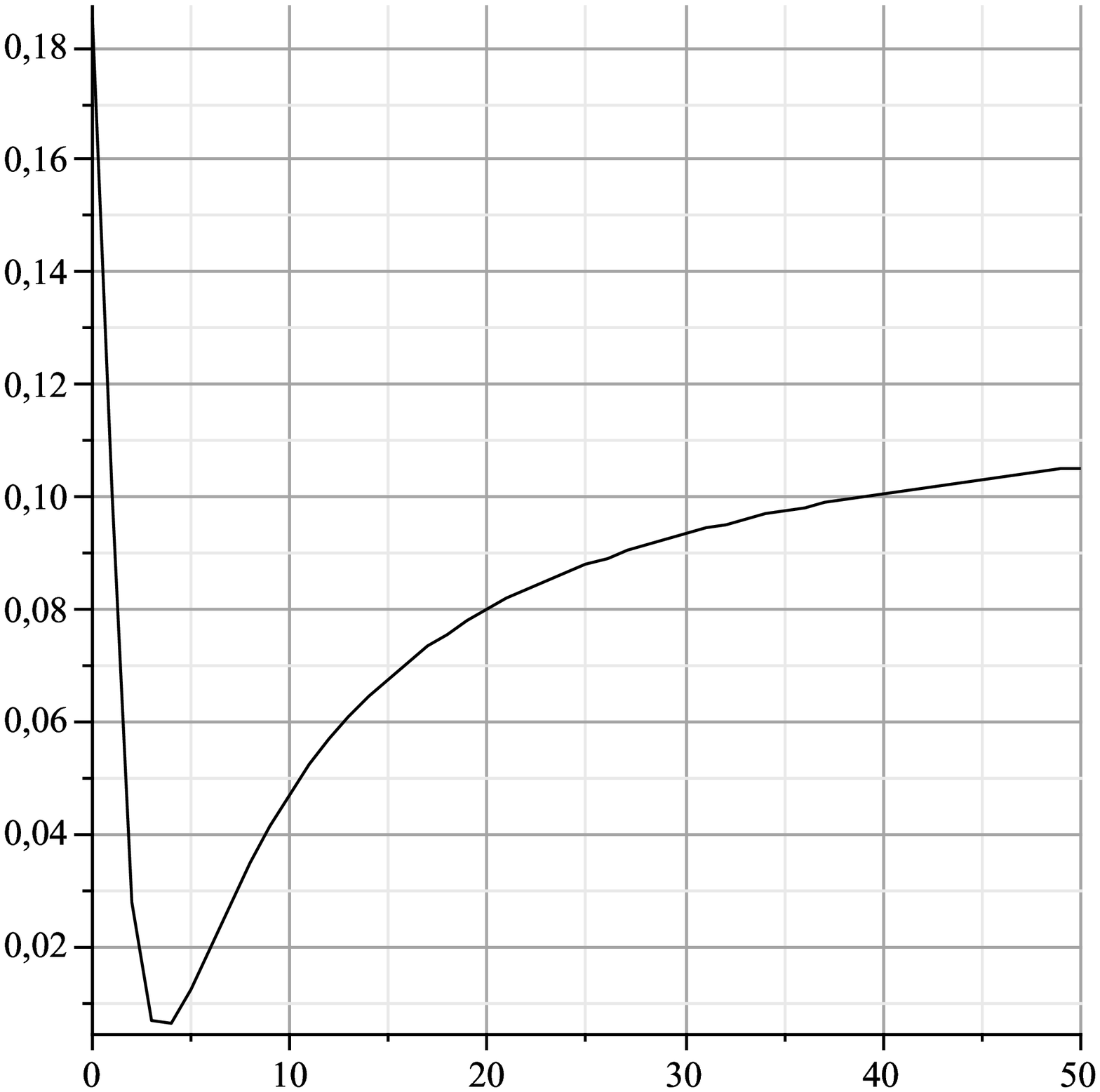}A.}
\put(65,-5){\includegraphics[width=7cm]{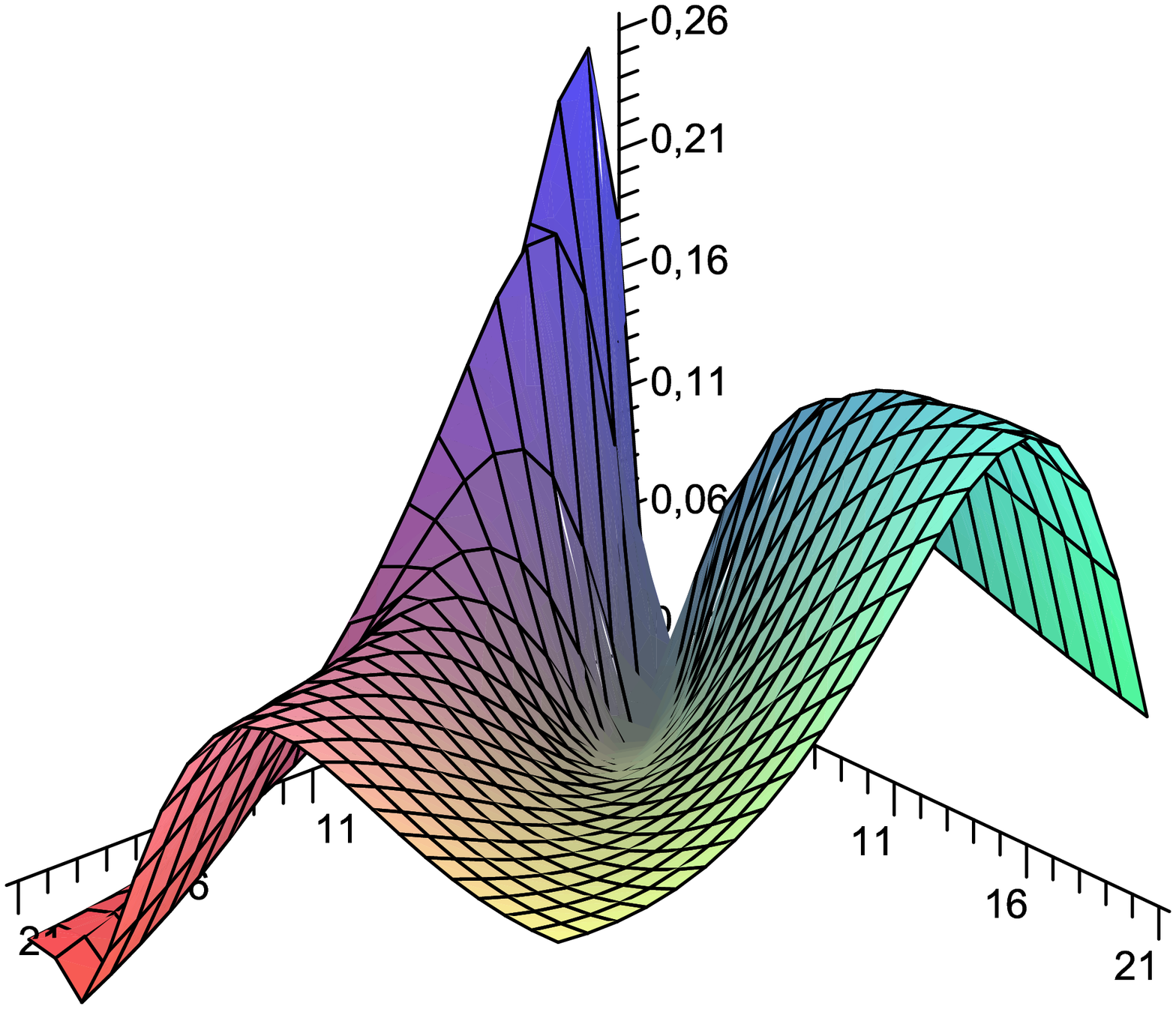}B.}
\put(25,-3){\makebox(0,0)[lb]{$N$}}
\put(-5,15){\begin{sideways} {$\mathcal R_B(\Phi_{B,N} | N,1)$} \end{sideways}}
\put(66,6){\makebox(0,0)[lb]{$K$}}
\put(132,4){\makebox(0,0)[lb]{$N$}}
\put(90,56){\makebox(0,0)[lb]{$\mathcal R_B(\Phi_{B,K} | N,1)$}}
\end{picture}
\end{center}
\caption{Contraction of the equation of motion for $\Phi_{B,N}(1)$
with $\Phi_{B,K}(1)$:\newline A. $N=K$ and $1\leq N\leq 50$;\newline B. $1\leq N\leq
20$, $1\leq K\leq 20$} \label{Rb2}
\end{figure}

For this purpose  we evaluate the correlators
\begin{equation}
\label{cor-pgc-pgc} \Rpg{B}{\Phi_{B,K}}{N}\equiv
\langle\Phi_{B,K}(\lambda),Q\Phi_{B,N}(\lambda)+\Phi_{B,N}(\lambda)\star
\Phi_{B,N}(\lambda)\rangle,
\end{equation}
and draw $\Rpg{B}{\Phi_{B,K}}{N}$ for $K=N$ as a function of $N$ for
$\lambda=1$. In figure  \ref{Rb2}.A we see that
$\Rpl{B}{\Phi_{B,N}}{N}$ does not go to zero when $N\to \infty$.

In figure  \ref{Rb2}.B we show the dependence of
$\Rpl{B}{\Phi_{B,K}}{N}$ on $K $ and $N$ and one can  see that
 there are directions along which $\Rpl{B}{\Phi_{B,K}}{N}$ does not go to zero when $N,K\to \infty$.

\subsection{Contractions  for the  Schnabl Solution}

Since the  pure gauge configuration doesn't solve equation of motion
at $\lambda=1$ one can try to add the phantom term. In this
subsection we will show numerically that this phantom term  minimize
deviations from the solution. We add the Schnabl phantom term to
the pure gauge solution with an arbitrary
 coefficient\footnote{We perform calculations in  \cite{Sch, FuchsKr} notations.
 To fit our general notations in section 2 we put the minus in front of the first term of (\ref{Phi^R})}

 \bea
 \Phi ^R_{B,N}(a)&=&-\Phi_{B,N}(1)+a\varphi_N\label{Phi^R}
 \eea
 and consider
 \be
\mathcal R_B^R(\varphi_K|N,1,a)=
\langle\varphi_{K},Q\Phi_{B,N}^{R}(a)+\Phi_{B,N}^{R}(a)\star
\Phi_{B,N}^{R}(a)\rangle. \ee

\begin{figure}[!h]
\begin {center}
\setlength{\unitlength}{1mm}
\begin{picture}(130,60)
\put(45,50){N=30}\put(52,33){N=40}\put(52,25){N=50}
\put(0,0){\includegraphics[width=5cm]{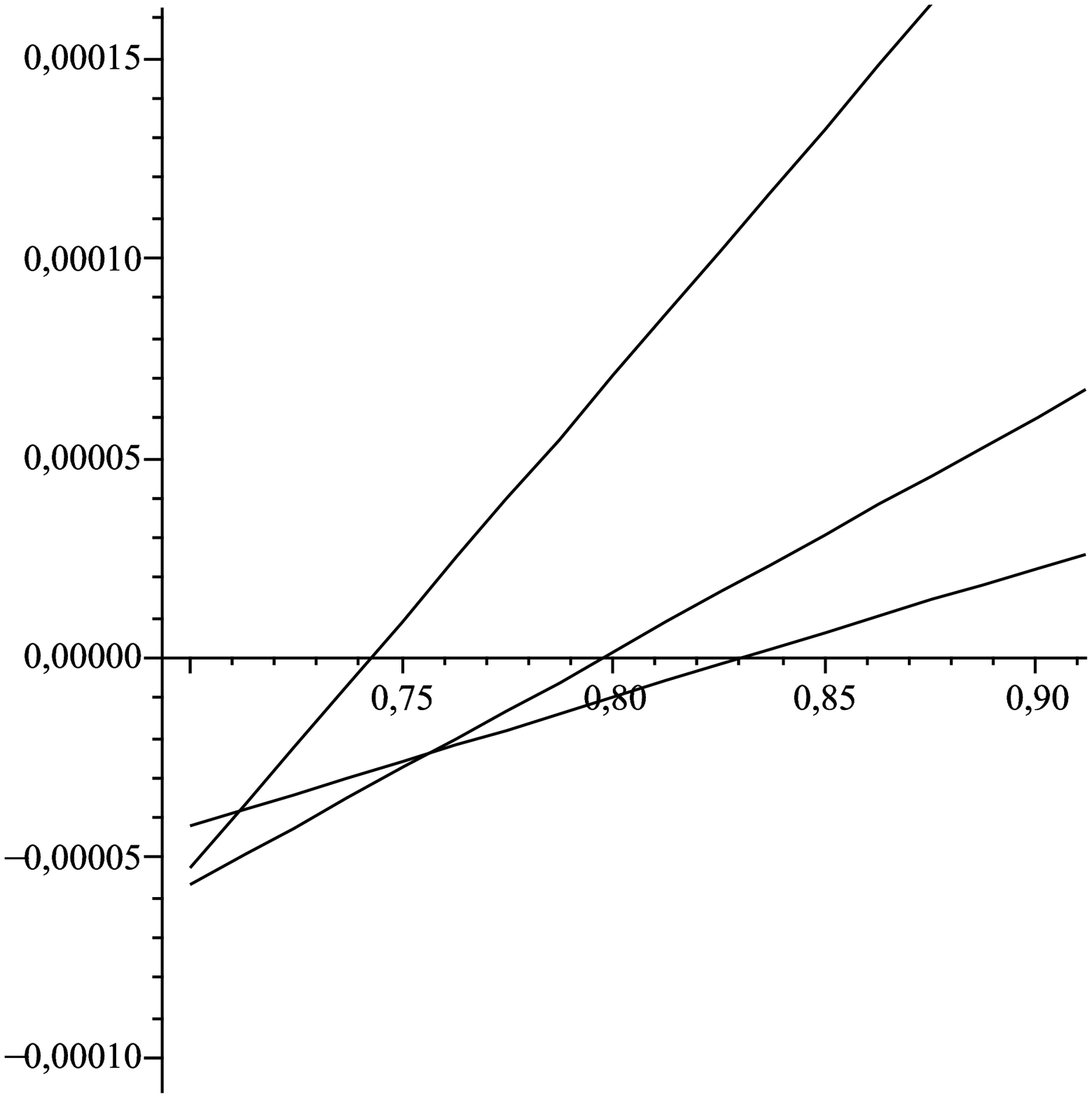}A.}
\put(75,0){\includegraphics[width=5cm]{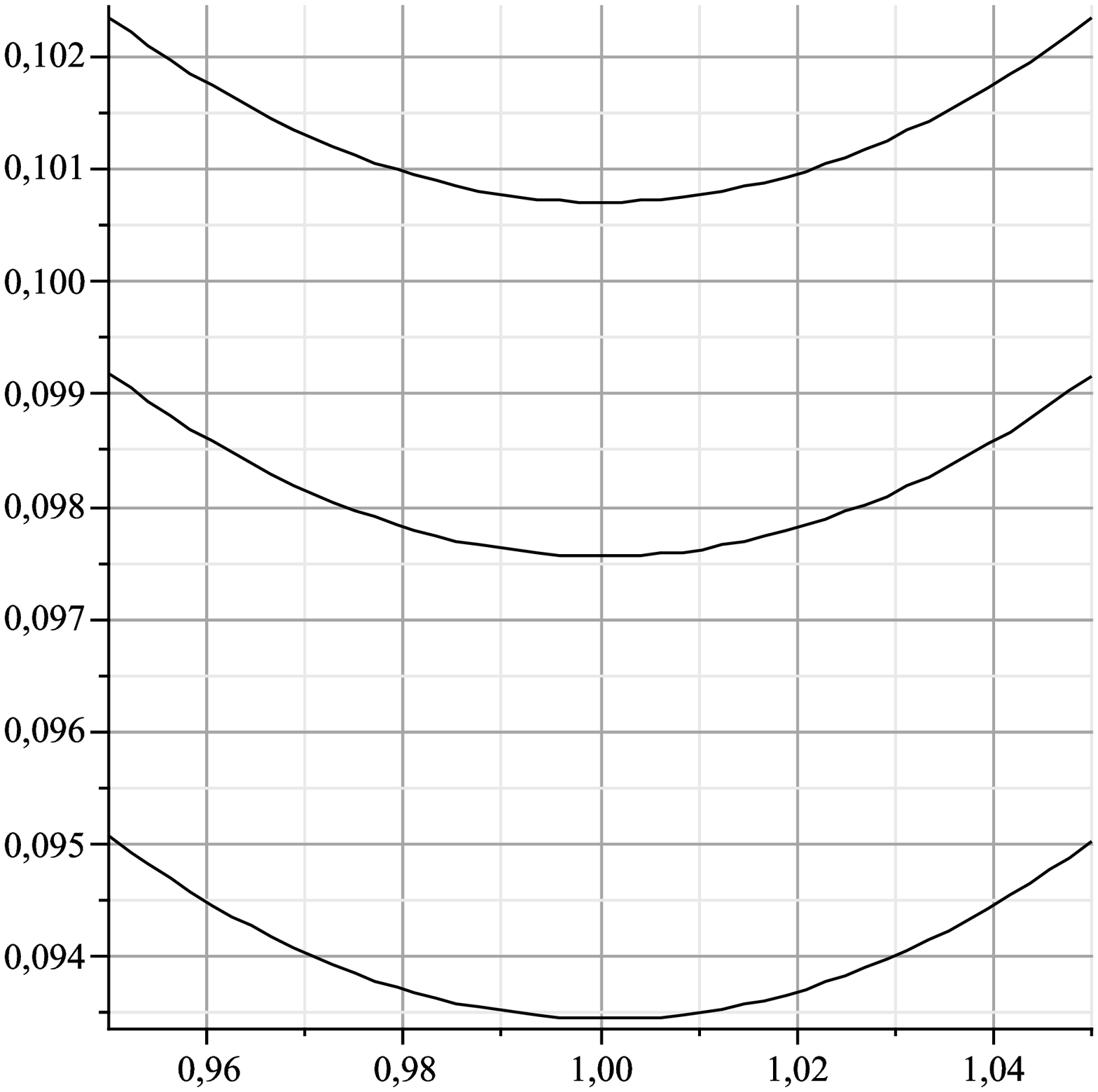}B.}
\put(30,15){\makebox(0,0)[lb]{$a$}}
\put(-4,10){\begin{sideways} {$\mathcal R_{N,B}^R(\varphi_1 |N,1,a)$} \end{sideways}}
\put(93,-3){\makebox(0,0)[lb]{$\lambda$}}
\put(68,8){\begin{sideways} {$\mathcal R_{N,B}^R(\Phi ^R_N(a)|N,1,a)$} \end{sideways}}
\put(127,50){N=30}\put(127,32){N=35}\put(127,12){N=40}\end{picture}
\end{center}
\caption{\newline A. Contraction of the
equation of motion for $\Phi^R_{B,N}(a)$ with $\varphi_1$,
as function of $a$ for different values of $N$, $N=30,40,50$. \newline B. $\mathcal R_{N,B}^R(\Phi ^R_N(a)|N,1,a)$
as a function of different
values of $N$,  $N=30,35,40$} \label{Rb5}
\end{figure}

In figure \ref{Rb5}.A we plot $\mathcal R_B^R(\varphi_1|N,1,a)$ for
different values of $a$ and $N=30,40,50$. We see that $\mathcal
R_B^R(\varphi_1|N,1,a)$ is equal to zero for these particular value
of $N$ for $a=a(N)$, which is very closed to $1$ and a deviation from
1 becomes smaller when
 $N$ increases.

 In figure \ref{Rb5}.B we plot $\mathcal R_B^R(\Phi ^R_N(a)|N,1,a)$ for different
values of $a$ and $N$. We see that $a=1$ minimize the deviation from
the  zero of $\mathcal R_B^R(\Phi ^R_N(a)|N,1,a)$ for particular
values of $N$ and this deviation decreases when $N$ increases.

\subsection{Calculation of Action on the Schnabl Solution}
It is known that the value of the action on the Schnabl solution
multiplied on $2\pi^2$ is equal to -1. This has been checked in
\cite{Sch, Okawa, FuchsKr} in the sense that \be
S=\lim_{N\to\infty}S(N)=-\frac{1}{2\pi^2}\label{SS},
\ee where
we use the following notations
\begin{equation}
S(N)\equiv S(\Phi^R_{B,N}(1)).\label{act}
\end{equation}

\begin{figure}[!h]
\begin {center}
\setlength{\unitlength}{1mm}
\begin{picture}(130,60)
\put(0,0){\includegraphics[width=5cm]{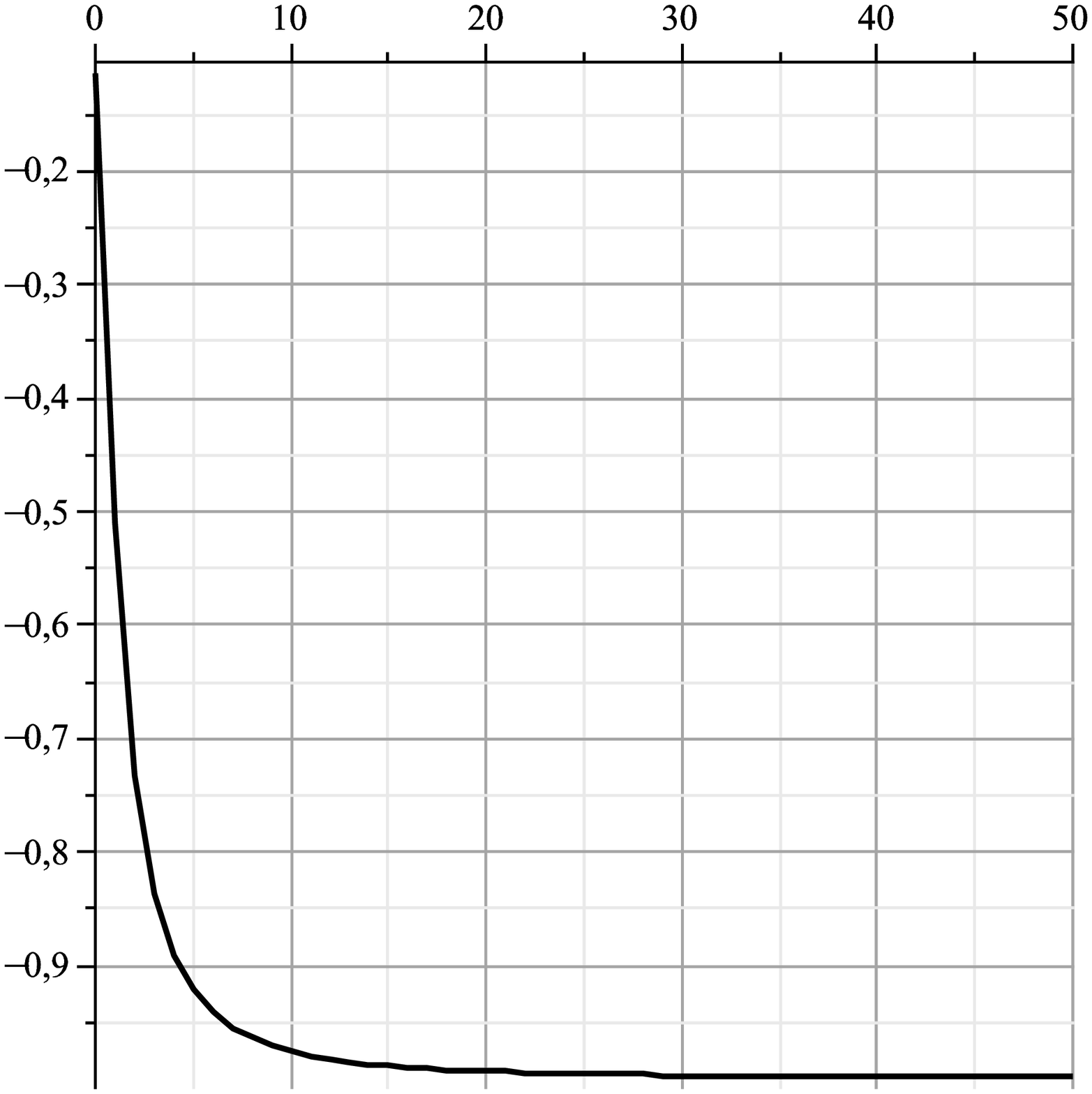}A.}
\put(65,0){\includegraphics[width=5cm]{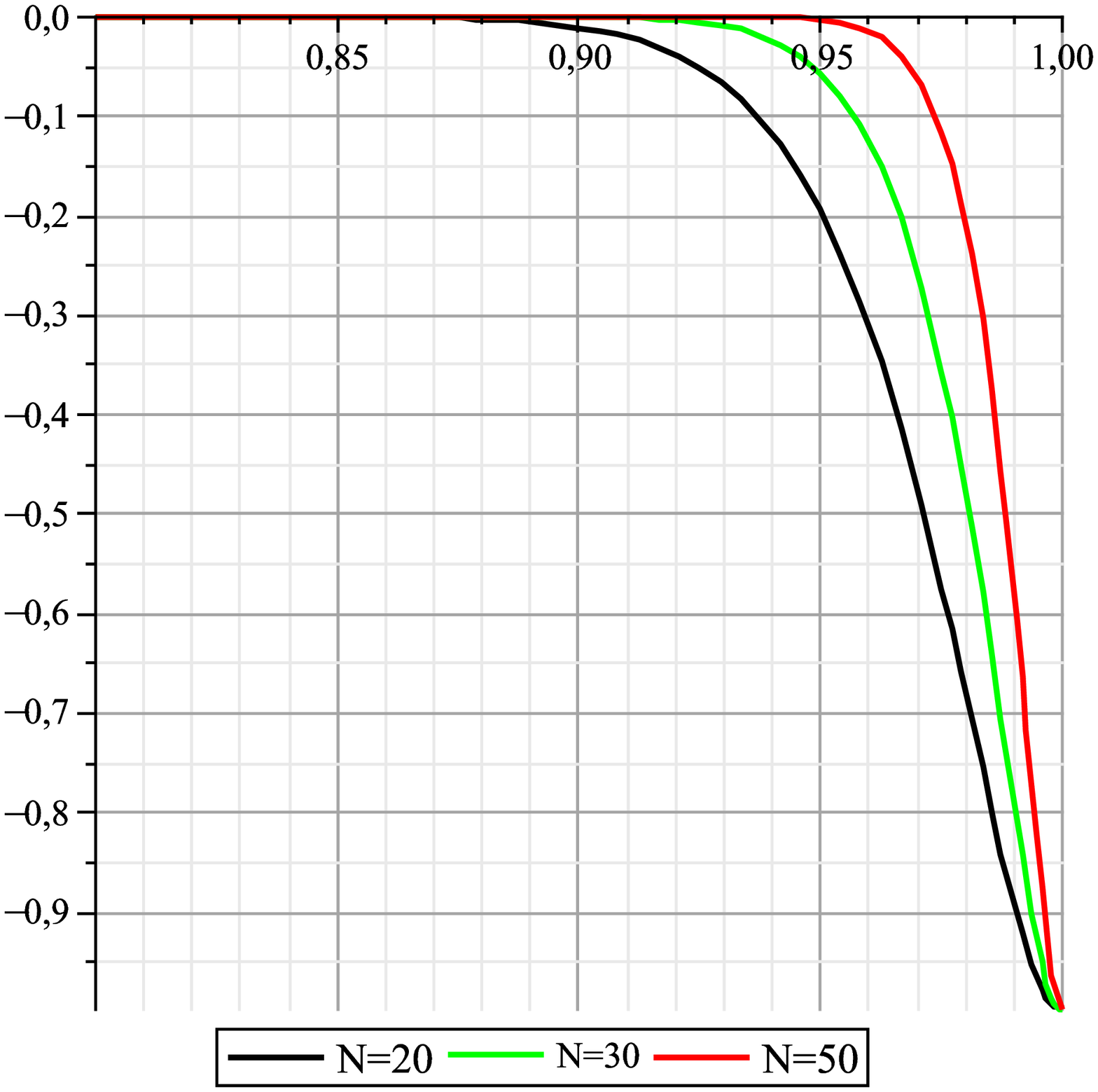}B.}
\put(25,50){\makebox(0,0)[lb]{$N$}} \put(-6,20){\begin{sideways}
{$2\pi^2S(\Phi^R_N(1 ))$} \end{sideways}}
\put(90,50){\makebox(0,0)[lb]{$\lambda$}}
\put(60,20){\begin{sideways} {$2\pi^2S(\Phi^R_N(\lambda ))$}
\end{sideways}}
\end{picture}
\end{center}
\caption{A. Value of the action multiplied by $2\pi^2$ on
$\Phi^R_N(\lambda,1) $ as function of $N$ for $\lambda=1$;
\,\,\,\,\,\,B.
 Value of the action multiplied by $2\pi^2$ on $\Phi^R_N(\lambda,1 )$ as function
 of  $\lambda$ for $N=20,30,50$} \label{Rb6}
\end{figure}

We calculate the action on partial sums (\ref{act}) for $a=1$ and
present the graph for $S(N)$ in figure \ref{Rb6}.A. We also calculate the action on
$\Phi^R_N(\lambda,1 )=-\Phi_{B,N}(\lambda)+\lambda^{N+1}\varphi_N$
and plot $\Phi^R_N(\lambda,1 )$ as function of  $\lambda$ for $N=20,30,50$
in figure \ref{Rb6}.B. We see that the values of  $2\pi^2\Phi^R_N(\lambda,1 )$
are closed  to $-1$ when $\lambda\to 1$ and $N$ is big enough.

\newpage
$$\,$$
\newpage
\section{Conclusion and Outlook}

In this paper we have studied the special class of pure gauge
configurations in the Witten bosonic SFT, the cubic SSFT  and the
fermionic SFT including the $GSO(-)$ sector. All these configurations are parameterized
by one parameter $\lambda$ and  are constructed as perturbation
expansions in $\lambda$. One can expect that these configurations
solve corresponding string field equations of motion. However since  they
are constructed as perturbation
expansions  the best  that we can expect is the validity of the
corresponding perturbative  expansions at each order of the
perturbation parameter. To find physical quantities related to
these configurations one  has to deal with these configurations as a
whole. Therefore one needs to study the convergence of the
perturbation expansions.

The simplest possibility  to deal with the convergence problem is to
consider existence of weak asymptotic solutions to equations of
motion in a sense of definition (\ref{asymp}). Already in this
simplest framework  we have seen  that for the large parameter of the
perturbation expansion the pure gauge truncated configurations give
divergent contributions to the equation of motion on the subspace of
the wedge states. In particular, on the example of the Erler pure gauge
configuration in the SSFT one can see explicitly (\ref{Erler-weak-EOM-pg})
that the perturbative pure gauge
configuration does not solve the equation of motion at $\lambda\geq1$.
We have also seen numerically the similar  effect for the pure gauge
configurations related to tachyon solutions for the bosonic and the
NS fermionic SFT. We have seen a difference in the behavior of
correlators of the equations of motion for string fields with wedge states
for the bosonic and fermionic cases.
For the bosonic case the equation of motion is not satisfied in the
{\it uniform} weak sense while for the fermionic  cases the equation of motion
is not satisfied already in the
weak sense at $\lambda=1$.

By  analytical calculations for the SSFT case and by the numerical
ones for the tachyon cases we  have shown  that the perturbation
expansions are cured by adding extra terms. In the tachyon cases
these terms  coincide with the terms found before in \cite{Sch} and \cite{AGM}
for the bosonic and fermionic string,
respectively, from requirements  to implement the
Sen conjectures. In the case of superstring these extra terms also coincide
with the Erler phantom terms and this justified the Erler choice of these terms
since a priori there is no reason to have a special number for the value of the
action on this solution.

All currently known analytical solutions can be cast
in a form of formal gauge solutions and one can hope that  all string field theory
solutions are of this form (see detailed discussion of this issue in the Fuchs
and Kroyter recent review \cite{KF-rev}).
A check of the equation of motion in a weak sense on wedge states could help to find   a  simple
prescription for regularizing formal solutions.

In particular, it is worth to study this problem for time depending rolling tachyon solutions
\footnote{For early constructions of  exact solutions in open bosonic string field
theory using marginal
  deformation in CFT \cite{SZmarg} see \cite{TT,Kluson1,Kluson2,super_id}}
that have been constructed perturbatively \cite{MS-TR,KORZ-mar,Okawa2,Erler2,
FKP,FK2}.
As has been found by Ellwood \cite{Ellwood1}  the late time behavior of the rolling tachyon solution
\cite{MS-TR,KORZ-mar}
approaches Schnabl's solution in the sense of correlators with  fields
belonging to the  Fock space. However the phantom term does not show up  in this consideration.

As has been noted in \cite{Erler} a similar limit for the
superstring \cite{Okawa2,Erler2}
fails to yield a well-defined expression and it would be interesting to study a
similar problem for the solution with nonvanishing $GSO(-)$ sector.

All pure gauge configurations are
gauge equivalent, but the singularity problem  and the necessity of adding the
extra terms means that the gauge equivalence  can be violated for
non-trivial solutions and this question needs of
a rather delicate study. It may turn out that it is not enough to calculate only the action to study gauge equivalence, so we have to consider other gauge invariants. Full list of gauge invariant quantities is unknown, but
it  includes the invariants related to the
1-point disk scattering amplitudes of closed
strings \cite{Hashimoto,GRSZ}. It would be interesting to clarify the question
about gauge equivalence of the Erler  and AGM solutions performing
1-point disk calculations.

Rolling tachyon solutions \cite{Sen-review,AJK} play important role in cosmology
\cite{AI,AJV,Calcagni,LJ-PR,Cline,AK-dil,BMNR}. These solutions exist in the flat background
 within the level truncation scheme
for the fermionic string \cite{YV,AJK,DP}, however for the bosonic string   wide
 oscillations do exist.
Moreover, there is the no-go theorem about existence of rolling solutions
for a toy model of the bosonic level truncated model, that is the $p$-adic string model for $p=2$ \cite{MZ}.  There are existence theorems
for $p=3$ $p$-adic string model \cite{VV} and similar models \cite{LJ-TMF}, which are the toy models of the
fermionic level truncated model with $GSO(-)$ sector. Note that a nontrivial background can change the situation for toy models  as well as for realistic models \cite{LJ-PR}. The wide oscillation  behavior is in the apparent conflict with the exponential grow result found using BSFT
and a resolution of this contradiction can be associated with the non-local field
redefinition between the two theories \cite{Coletti}.
A direct evaluation of the partition
function of the rolling tachyon solution of~\cite{MS-TR,KORZ-mar},
gives a result very similar to the one obtained in BSFT~\cite{Jokela}.
From cosmological perspectives results \cite{LJ-PR} could means that the field
redefinition becomes more smooth in the FRW background.
Note also cosmological applications \cite{AK-dil,BMNR} of the Hellerman and Schnabl
light-like rolling
solutions in SFT \cite{Hellerman-MS}.

\newpage

\section*{Acknowledgements}
The work is supported in part by RFBR grant 08-01-00798 and grant
NS-795.2008.1.

\section{Appendix}
\appendix

\section{Split-Strings Formalism}

Our split-string notations are based on Okawa's paper \cite{Okawa}.
A state written in the split-string notation can be rewritten in
the conformal language. For example:
\begin{equation}
\begin{split}
Fc\Omega^nBcF &= c(0)\ket{0}\star\ket{n}\star B_1^Lc(0)\ket{0}\\
&= \frac{1}{\pi}U_{n+2}^{\dagger}U_{n+2}\left[ \left(\mathcal B_0+
\mathcal B_0^{\dagger}\right)\tilde c\left(\frac{\pi}{4}n\right)
\tilde c\left(-\frac{\pi}{4}n\right) -\frac{\pi}{2}
\left(\tilde c\left(\frac{\pi}{4}n\right)+\tilde c
\left(-\frac{\pi}{4}n\right)\right)\right]\ket{0}.
\end{split}
\end{equation}
One uses  split-string notations to calculate correlators on
a cylinder of some circumference and these correlators  can be reduced to the
correlators on the unit disc by a suitable conformal transformation. This
provides the connection between the split-string formalism and the familiar
conformal language.

Further we collect some notations and useful formulae.
We use following string fields
\begin{equation}
\begin{split}
K &= K_1^L\ket{I}\quad\mathrm{Grassmann\ even,\ gh}\#= 0,\\
B &= B_1^L\ket{I}\quad\mathrm{Grassmann\ odd,\ gh}\#= -1,\\
c &= c(0)\ket{I}\quad\mathrm{Grassmann\ odd,\ gh}\#= 1,\\
\gamma &= \gamma(0)\ket{I}\quad\mathrm{Grassmann\ even,\ gh}\#=1,\\
\gamma^2 &= \gamma^2(0)\ket{I}\quad\mathrm{Grassmann\ even,\ gh}\#= 2,
\end{split}
\end{equation}
which satisfy the following algebraic relations:
\begin{equation}\label{eq:SSid}
\begin{split}
\{B,c\} &= 1,\quad [K,B] = 0,\quad B^2 = c^2= 0,\\
[B,\gamma] &= 0,\quad [c,\gamma]=0,\\
dK &= 0,\quad dB = K,\\
dc &= cKc-\gamma^2,\\
d\gamma &= cK\gamma-\frac12\gamma Kc-\frac12\gamma cK,\\
d\gamma^2 &= cK\gamma^2-\gamma^2 Kc,
\end{split}
\end{equation}
where $d=Q_B$ is the BRST operator.

We also use
\begin{equation}
F=e^{\frac\pi4K}=\Omega^{\frac12},
\end{equation}
which is the square root of the $SL(2,\mathbb{R})$ vacuum
$\Omega=e^{\frac\pi2K}$

\section{Correlators}
Using the split-string formalism we obtain the following correlators
\subsection{Quadratic correlators}
\begin{equation}\label{app-quadratic-correlators}
\begin{split}
\cors{\psi_n,Q\psi_k} &= \frac{n+k+2}{\pi^2},\\
\cors{\chi_n,Q\chi_k} &= 0,\\
\cors{\psi_n,Q\chi_k} &= \cors{\chi_k,Q\psi_n}\\
    &=-\frac{1}{\pi^3}\left[(k+2)\cos\ar{\pi k}{n+k+2} -\frac{\pi nk}{n+k+2}\sin\ar{\pi k}{n+k+2}\right],\\
\cors{\vartheta_n,Q\vartheta_k} &= \cors{\eta_n,Q\eta_k}\\
    &=-\frac{1}{\pi^3}\left[ \frac{k-n}{2}\cos\ar{\pi(n+1)}{n+k+2} +\frac{\pi(nk-1)}{n+k+2}\sin\ar{\pi(n+1)}{n+k+2} \right],\\
\cors{\vartheta_n,Q\eta_k} &= \cors{\vartheta_k,Q\eta_n} = \cors{\eta_n,Q\vartheta_k} = \cors{\eta_k,Q\vartheta_n}\\
    &=-\frac{1}{\pi^3}\left[ \cos\ar{\pi}{n+k+2} -\frac{\pi(n+k+1)}{n+k+2}\sin\ar{\pi}{n+k+2} \right].\\
\end{split}
\end{equation}

\subsection{Cubic correlators}

Correlators without 0-th term:
\begin{equation}\label{app-cubic-correlators-0}
\begin{split}
\cors{\psi_n,\psi_m,\chi_k} &= -\frac{n+m+k+3}{\pi^3}\cos\ar{\pi k}{n+m+k+3},\\
\cors{\psi_n,\vartheta_m,\vartheta_k} &=
    -\frac{n+m+k+3}{\pi^3}\cos\ar{\pi(m+1)}{n+m+k+3},\\
\cors{\psi_n,\vartheta_m,\eta_k} &=
    \frac{n+m+k+3}{\pi^3}\cos\ar{\pi(n+2)}{n+m+k+3},\\
\cors{\psi_n,\eta_m,\eta_k} &=
    -\frac{n+m+k+3}{\pi^3}\cos\ar{\pi(k+1)}{n+m+k+3}.\\
\end{split}
\end{equation}
Correlators with one 0-th term:
\begin{equation}\label{app-cubic-correlators-1}
\begin{split}
\cors{\zeta_0',\psi_n,\psi_k} &= -\frac{n+k+3}{2\pi^2},\\
\cors{\zeta_0',\psi_n,\chi_k} &= \cors{\zeta_0',\chi_k,\psi_n}
    = -\frac{1}{\pi^3}\left[\cos\ar{\pi k}{n+k+3} +\frac{\pi k}{n+k+3}\sin\ar{\pi k}{n+k+3}\right],\\
\cors{\zeta_0',\chi_n,\chi_k} &= 0,\\
\cors{\xi_0',\psi_n,\eta_k} &= -\cors{\xi_0',\vartheta_n,\psi_k}\\
    &= \frac{1}{\pi^3}\left[\cos\ar{\pi}{n+k+3} -\frac{\pi(n+k+2)}{n+k+3}\sin\ar{\pi}{n+k+3}\right],\\
\cors{\xi_0',\psi_n,\vartheta_k} &= -\cors{\xi_0',\eta_k,\psi_n}\\
    &=-\frac{1}{\pi^3}\left[ \frac{n+k+7}{2}\cos\ar{\pi(n+2)}{n+k+3}
    +\frac{\pi(n-k+1)}{n+k+3}\sin\ar{\pi(n+2)}{n+k+3}\right],\\
\cors{\zeta'_0,\vartheta_n,\vartheta_k} &=
    -\frac{1}{\pi^3}\left[\cos\ar{\pi(n+1)}{n+k+3} +\frac{\pi(n+1)}{n+k+3}\sin\ar{\pi(n+1)}{n+k+3}\right],\\
\cors{\zeta'_0,\vartheta_n,\eta_k} &=
    \frac{1}{\pi^3}\left[\cos\ar{2\pi}{n+k+3} -\frac{\pi(n+k+1)}{n+k+3}\sin\ar{2\pi}{n+k+3}\right],\\
\cors{\zeta'_0,\eta_n,\vartheta_k} &= 0,\\
\cors{\zeta'_0,\eta_n,\eta_k} &=
    -\frac{1}{\pi^3}\left[\cos\ar{\pi(k+1)}{n+k+3} +\frac{\pi(k+1)}{n+k+3}\sin\ar{\pi(k+1)}{n+k+3}\right].\\
\end{split}
\end{equation}
Correlators with two 0-th terms:
\begin{equation}\label{app-cubic-correlators-2}
\begin{split}
\cors{\zeta_0',\zeta_0',\psi_n} &= -\frac{1}{\pi^2},\\
\cors{\zeta_0',\zeta_0',\chi_n} &=
    \frac{n^2}{\pi (n+3)^3}\cos\ar{\pi n}{n+3},\\
\cors{\zeta_0',\xi_0',\vartheta_n} &= -\cors{\xi_0',\zeta_0',\eta_n}
    =-\frac{n+2}{\pi (n+3)^3}\cos\ar{\pi}{n+3},\\
\cors{\zeta_0',\xi_0',\eta_n} &=
    +\frac{1}{\pi^3}\left[\left(\frac12 +\frac{\pi^2(1-n^2)}{(n+3)^3} \right)\cos\ar{2\pi}{n+3}
    -\frac{\pi(n+1)}{2(n+3)}\sin\ar{2\pi}{n+3}\right], \\
\cors{\xi_0',\zeta_0',\vartheta_n} &=
    -\frac{1}{\pi^3}\left[ \left( \frac 12 +\frac{\pi^2(1-n^2)}{(n+3)^3} \right)\cos\ar{2\pi}{n+3}
    -\frac{\pi(n+1)}{2(n+3)}\sin\ar{2\pi}{n+3}\right],\\
\cors{\xi_0',\xi_0',\psi_n} &=
    \frac{1}{\pi^3}\left[ \left(-1+\frac{\pi^2n(n+2)}{(n+3)^3}\right)\cos\ar{\pi}{n+3}
    +\frac{\pi(n+2)}{n+3}\sin\ar{\pi}{n+3}\right].\\
\end{split}
\end{equation}


\newpage
\begin{thebibliography}{72}
\bibitem{Sch}
M.~Schnabl, {\it Analytic solution for tachyon condensation in open
string field theory}, \newjournal{Adv.\ Theor.\
Math.\ Phys.\ }{TMPHA}{10}{2006}{433},
[\hepth{0511286}].

\bibitem{W}
E.~Witten, {\it Interacting field theory of open superstrings}, \npb{276}{1986}{291}.


\bibitem{Okawa}
  Y.~Okawa,
  {\it Comments on Schnabl's analytic solution for tachyon condensation in
  Witten's open string field theory,}
  \jhep{0604}{055}{2006}
  [\hepth{0603159}].

\bibitem{FuchsKr} E. Fuchs and M. Kroyter, {\it On the validity of the solution
of string field theory,}
 \jhep{0605}{2006}{006}, [\hepth{0603195}].

\bibitem{RZ}
  L.~Rastelli and B.~Zwiebach,
  {\it Solving open string field theory with special projectors,}
  \jhep{0801}{020}{2008}
  [\hepth{0606131}].

\bibitem{ORZ}
  Y.~Okawa, L.~Rastelli and B.~Zwiebach,
  {\it Analytic solutions for tachyon condensation with general projectors,}
  [\hepth{0611110}].

\bibitem{MS-TR}  M.~Schnabl, {\it Comments on Marginal Deformations
 in Open String Field Theory,} [\hepth{0701248}]


\bibitem{KORZ-mar}  M. Kiermaier,
Y.Okawa, L.Rastelli and B.Zwiebach, {\it Analytic Solutions for
Marginal Deformations in Open String Field Theory,}
[\hepth{0701249}].

\bibitem{RZ2}
  L.~Rastelli and B.~Zwiebach,
  {\it The off-shell Veneziano amplitude in Schnabl gauge,}
  \jhep{0801}{018}{2008}
  [\hepth{0708.2591}].

\bibitem{KSZ}
  M.~Kiermaier, A.~Sen and B.~Zwiebach,
  {\it Linear b-Gauges for Open String Fields,}
  \jhep{0803}{050}{2008}
  [\hepth{0712.0627}].

\bibitem{KZ}
  M.~Kiermaier and B.~Zwiebach,
  {\it One-Loop Riemann Surfaces in Schnabl Gauge,}
  \jhep{0807}{063}{2008}
  [\hepth{0805.3701}].

\bibitem{KOZ}
  M.~Kiermaier, Y.~Okawa and B.~Zwiebach,
  {\it The boundary state from open string fields,}
  [\hepth{0810.1737}].

\bibitem{ES}
  I.~Ellwood and M.~Schnabl,
  {\it Proof of vanishing cohomology at the tachyon vacuum,}
  \jhep{0702}{096}{2007}
  [\hepth{0606142}].

\bibitem{Ellwood1}
  I.~Ellwood,
  {\it Rolling to the tachyon vacuum in string field theory,}
  \jhep{0712}{028}{2007}
  [\hepth{0705.0013}].\\
\bibitem{Ellwood2}  I.~Ellwood, {\it The closed string tadpole in open string field theory,}
  \jhep{0808}{063}{2008}
  [\hepth{0804.1131}].

\bibitem{FK3}
  E.~Fuchs and M.~Kroyter,
  {\it Schnabl's $L_0$ operator in the continuous basis,}
  \jhep{0610}{067}{2006}
  [\hepth{0605254}].

\bibitem{FKP}
  E.~Fuchs, M.~Kroyter and R.~Potting,
  {\it Marginal deformations in string field theory,}
  \jhep{0709}{101}{2007}
  [\hepth{0704.2222}].

\bibitem{FK2}
  E.~Fuchs and M.~Kroyter,
  {\it Marginal deformation for the photon in superstring field theory,}
  \jhep{0711}{005}{2007}
  [\hepth{0706.0717}].

\bibitem{Erler3}
  T.~Erler,
  {\it Split string formalism and the closed string vacuum,}
  \jhep{0705}{083}{2007}
  [\hepth{0611200}].\\
T.~Erler,
  {\it Split string formalism and the closed string vacuum. II,}
  \jhep{0705}{084}{2007}
  [\hepth{0612050}].

\bibitem{BMST}
  L.~Bonora, C.~Maccaferri, R.~J.~Scherer Santos and D.~D.~Tolla,
  {\it Ghost story. I. Wedge states in the oscillator formalism,}
  \jhep{0709}{061}{2007}
  [\hepth{0706.1025}].

\bibitem{T}
  T.~Takahashi,
  {\it Level truncation analysis of exact solutions in open string field theory,}
  \jhep{0801}{001}{2008}
  [\hepth{0710.5358}].

\bibitem{KKT2}
  T.~Kawano, I.~Kishimoto and T.~Takahashi,
  {\it Gauge Invariant Overlaps for Classical Solutions in Open String Field
  Theory,}
  \npb{803}{135}{2008}
  [\hepth{0804.1541}].

\bibitem{KKT}
  T.~Kawano, I.~Kishimoto and T.~Takahashi,
  {\it Schnabl's Solution and Boundary States in Open String Field Theory,}
  \plb{669}{357}{2008}
  [\hepth{0804.4414}].

\bibitem{Kwon:2007mh}
  O.~K.~Kwon, B.~H.~Lee, C.~Park and S.~J.~Sin,
  {\it Fluctuations around the Tachyon Vacuum in Open String Field Theory,}
  \jhep{0712}{038}{2007}
  [\hepth{0709.2888}].

\bibitem{Kwon:2008ap}
  O.~K.~Kwon,
  {\it Marginally Deformed Rolling Tachyon around the Tachyon Vacuum in Open
  String Field Theory,}
  \npb{804}{1}{2008}
  [\hepth{0801.0573}].

\bibitem{Ishida:2008jc}
  A.~Ishida, C.~Kim, Y.~Kim, O.~K.~Kwon and D.~D.~Tolla,
  {\it Tachyon Vacuum Solution in Open String Field Theory with Constant B
  Field,}
  [\hepth{0804.4380}].

\bibitem{Erler2}
  T.~Erler,
  {\it Marginal Solutions for the Superstring,}
  \jhep{0707}{050}{2007}
  [\hepth{0704.0930}].

  \bibitem{Okawa2}
  Y.~Okawa,
  {\it Analytic solutions for marginal deformations in open superstring field
  theory,}
  \jhep{0709}{084}{2007}
  [\hepth{0704.0936}].\\
  Y.~Okawa,
  {\it Real analytic solutions for marginal deformations in open superstring field
  theory,}
  \jhep{0709}{082}{2007}
  [\hepth{0704.3612}].\\
  M.~Kiermaier and Y.~Okawa,
  {\it General marginal deformations in open superstring field theory,}
  [\hepth{0708.3394}].



\bibitem{Erler}
  T.~Erler,
  {\it Tachyon Vacuum in Cubic Superstring Field Theory,}
  \jhep{0801}{013}{2008}
  [\hepth{0707.4591}].

\bibitem{AGM}
  I.~Y.~Aref'eva, R.~V.~Gorbachev and P.~B.~Medvedev,
  {\it Tachyon Solution in Cubic Neveu-Schwarz String Field Theory,}
  \newjournal{Theor.\ and Math.\ Phys.}{TMP}{158(3)}{320}{2009}
  [\hepth{0804.2017}].

  \bibitem{KF}
  E.~Fuchs and M.~Kroyter,
  {\it On the classical equivalence of superstring field theories,}
  [\hepth{0805.4386}].

   \bibitem{KF-rev} E.~Fuchs and M.~Kroyter,
  {\it Analytical Solutions of Open String Field Theory,}
  [\hepth{0807.4722}].



\bibitem{AGM-c}
    I.~Y.~Aref'eva, R.~V.~Gorbachev and P.~B.~Medvedev,
    {\it Vacuum solutions NS fermionic SFT and zero curvature representation in graded spaces,} submitted to Theor. Math. Phys. \\
    I.~Y.~Aref'eva, R.~V.~Gorbachev and P.~B.~Medvedev,
    {\it Pure Gauge Configurations and Solutions to Fermionic Superstring Field Theories Equations of Motion.}
    Talk presented at {\it Liouville Gravity and Statistical Models,} International conference
    in memory of Alexei Zamolodchikov, Moscow, June 21-24, 2008


\bibitem{AGMM}
    I.~Y.~Aref'eva, R.~V.~Gorbachev, M.Maltsev and P.~B.~Medvedev,
    {\it On Schnabl solutions to OSFT,}
    submitted to Proceedings of Steklov Math.Institute.





\bibitem{Sen}
 A.~Sen,
  {\it Stable non-BPS bound states of BPS D-branes,}
  \jhep{9808}{010}{1998}
  [\hepth{9805019}].

  A.~Sen,
  {\it SO(32) spinors of type I and other solitons on brane-antibrane pair,}
  \jhep{9809}{023}{1998}
  [\hepth{9808141}].


\bibitem{AMZ}
  I.~Y.~Arefeva, P.~B.~Medvedev and A.~P.~Zubarev,
  {\it Background Formalism For Superstring Field Theory,}
  \plb{240}{356}{1990}.

I.~Y.~Arefeva, P.~B.~Medvedev, and A.~P.~Zubarev, {\it New
representation for string field solves the consistence problem for
open superstring field,} \npb{341}{1990}{464}.

\bibitem{PTY}
C.~R.~Preitschopf, C.~B.~Thorn, and S.~A.~Yost, {\it Superstring field
theory,} \npb{337}{1990}{363}.

\bibitem{AMZ2}
  I.~Y.~Arefeva, P.~B.~Medvedev and A.~P.~Zubarev,
  {\it Nonperturbative vacuum for superstring field theory and supersymmetry
  breaking,}
  \newjournal{Mod.\ Phys.\ Lett.}{MPL}{A6}{949}{1991}.

\bibitem{GSW}
  M.~B.~Green, J.~H.~Schwarz and E.~Witten,
  {\it Superstring Theory. vol.~1: Introduction,}
{\it  Cambridge, UK: Univ. Pr. (1987) 469 P.}

\bibitem{ABKM}
   I.~Y.~Aref'eva, D.~M.~Belov, A.~S.~Koshelev and P.~B.~Medvedev,
  {\it Tachyon condensation in cubic superstring field theory,}
  \npb{638}{3}{2002}
  [\hepth{0011117}].

  I.~Y.~Arefeva, D.~M.~Belov, A.~S.~Koshelev and P.~B.~Medvedev,
  {\it Gauge invariance and tachyon condensation in cubic superstring field
  theory,}
  \npb{638}{21}{2002}
  [\hepth{0107197}].

\bibitem{ABGKM}
  I.~Y.~Arefeva, D.~M.~Belov, A.~A.~Giryavets, A.~S.~Koshelev and P.~B.~Medvedev,
  {\it Noncommutative field theories and (super)string field theories,}
  [\hepth{0111208}].


\bibitem{ABG}
I.~Y.~Arefeva, D.~M.~Belov, and A.~A.~Giryavets, {\it Construction of
the vacuum string field theory on a non-BPS brane,} \jhep{09}{2002}{050}, [\hepth{0201197}].

\bibitem{BSZ}
N.~Berkovits, A.~Sen, and B.~Zwiebach, {\it Tachyon condensation in
superstring field theory,} \npb{587}{2000}{147},
[\hepth{0002211}].

\bibitem{B}
N.~Berkovits, {\it Super-Poincare invariant superstring field theory},
\npb{450}{1995}{90}, [\hepth{9503099}].

\bibitem{FMS} D.~Friedan, E.~Martinec and
S.~Shenker, {\it Conformal Invariance, Supersymmetry and String
Theory}, \npb{271}{1986}{93}.

\bibitem{BSZ}
N.~Berkovits, A.~Sen, and B.~Zwiebach, {\it Tachyon condensation in
superstring field theory}, \npb{587}{2000}{147},
[\hepth{0002211}].

\bibitem{Rastelli:2000iu}
  L.~Rastelli and B.~Zwiebach,
 {\it Tachyon potentials, star products and universality,}
  \jhep{0109}{038}{2001}
  [\hepth{0006240}].

\bibitem{Rastelli:2001vb}
  L.~Rastelli, A.~Sen and B.~Zwiebach,
  {\it Boundary CFT construction of D-branes in vacuum string field theory,}
  \jhep{0111}{045}{2001}
  [\hepth{0105168}].

\bibitem{Schnabl2}
  M.~Schnabl,
  {\it Wedge states in string field theory,}
  \jhep{0301}{004}{2003}
  [\hepth{0201095}].



\bibitem{SZmarg}
  A.~Sen and B.~Zwiebach,
  {\it Large marginal deformations in string field theory,}
  \jhep{0010}{009}{2000}
  [\hepth{0007153}].

\bibitem{TT}
  T.~Takahashi and S.~Tanimoto,
  {\it Marginal and scalar solutions in cubic open string field theory,}
  \jhep{0203}{033}{2002}
  [\hepth{0202133}].

\bibitem{Kluson1}
  J.~Kluson,
  {\it Exact solutions in open bosonic string field theory and marginal
  deformation in CFT,}
  \newjournal{Int.\ J.\ Mod.\ Phys.}{IJMP}{A19}{4695}{2004}
  [\hepth{0209255}].

\bibitem{Kluson2}
  J.~Kluson,
  {\it Exact solutions in SFT and marginal deformation in BCFT,}
  \jhep{0312}{050}{2003}
  [\hepth{0303199}].
  \bibitem{super_id} I. Kishimoto and T. Takahashi, {\it Marginal
deformations and classical solutions in open superstring field theory,}
\jhep{0511}{051}{2005} [\hepth{0506240}].
\bibitem{Hashimoto}
A.~Hashimoto and N.~Itzhaki, {\it Observables of string field theory},
  \jhep{01}{028}{2002}
 [\hepth{0111092}].

  \bibitem{GRSZ}
D.~Gaiotto, L.~Rastelli, A.~Sen, and B.~Zwiebach, {\it Ghost structure and
  closed strings in vacuum string field theory},
 [\hepth{0111129}].

  \bibitem{MZ}
  N.~Moeller and B.~Zwiebach,
  {\it Dynamics with infinitely many time derivatives and rolling tachyons,}
  \jhep{0210}{034}{2002},
  [\hepth{0207107}].
\bibitem{Sen-review}
  A.~Sen,
  {\it Tachyon dynamics in open string theory,}
  \newjournal{Int.\ J.\ Mod.\ Phys.}{IJMP}{A20}{2005}{5513},
  [\hepth{0410103}].
  \bibitem{Coletti}
E.~Coletti, I.~Sigalov, and W.~Taylor, {\it Taming the tachyon in cubic string
  field theory},  \jhep{08}{2005}{104},
  [\hepth{0505031}].


  \bibitem{Jokela}
N.~Jokela, M.~Jarvinen, E.~Keski-Vakkuri, and J.~Majumder, {\it Disk partition
  function and oscillatory rolling tachyons},  \newjournal{J.\ Phys.}{JP}{A41}{2008}{015402},  [\hepth{0705.1916}].

\bibitem{AJK}  I.Ya.Aref'eva, L.V.Joukovskaya, A.S.Koshelev,
{\it Time evolution in superstring field theory on nonBPS brane. 1. Rolling tachyon and energy momentum conservation},
 \jhep{0309}{012}{2003}, [\hepth{0301137}]
\bibitem{YV} Ya. Volovich, {\it Numerical study of nonlinear equations with infinite
number of derivatives.}
\newjournal{J.\ Phys.}{JP}{A36}{8685}{2003}
[\arXivid{0301028}]
\bibitem{VV} V.S. Vladimirov, Y.I. Volovich,  {\it On the nonlinear dynamical equation in the p-adic string theory},
\newjournal{Theor.\ Math.\ Phys.}{TMP}{138}{297}{2004},
[\arXivid{0306018}]
\bibitem{LJ-TMF} L.V. Joukovskaya, {\it Iterative method for solving nonlinear
integral equations describing rolling solutions in string theory},
\newjournal{Theor.\ Math.\ Phys.}{TMP}{146}{335}{2006},
Teor.Mat.Fiz.146:402-409,2006, [\hepth{0708.0642}].
\bibitem{DP} D.V. Prokhorenko,  {\it On some nonlinear integral equation in the
(super)string theory},
[\arXivid{0611068}]
\bibitem{AI} I. Aref'eva,  {\it Nonlocal string tachyon as a model for
cosmological dark energy}.
AIP Conf. Proc. 826 301-311 (2006),  [\astroph{0410443}];  I.Ya. Aref'eva, Stringy Model of Cosmological Dark Energy, AIP Conf. Proc. 957 (2007)
297, [\hepth{0710.3017}]
\bibitem{AJV} I.Ya. Aref'eva, L.V. Joukovskaya, S.Yu. Vernov, {\it Bouncing
and accelerating solutions in nonlocal stringy models}, \jhep{0707}{087}{2007}
[\hepth{0701184}]; \newjournal{J.\ Phys.}{JP}{A41}{304003}{2008}, {\it  Dynamics in nonlocal linear models in the Friedmann-Robertson-Walker metric},  [\hepth{0711.1364}].
  \bibitem{Calcagni} G. Calcagni, {\it Cosmological Tachyon from Cubic String Field Theory}, \jhep{0605}{2006}{012},
  [\hepth{0512259}]; G. Calcagni, M. Montobbio, and G. Nardelli, A route to nonlocal cosmology,
  \prd{76}{2007}{126001}, [\hepth{0705.3043}]
\bibitem{Cline} N. Barnaby, T. Biswas and J.M. Cline, p-adic Inflation, \jhep{0704}{2007}{056},
[\hepth{0612230}];
N. Barnaby and J.M. Cline, {\it Large Nongaussianity from Nonlocal Inflation,} \newjournal{JCAP}{JCAP}{0707}{2007}{017}, [\hepth{0704.3426}]
\bibitem{LJ-PR} L.V. Joukovskaya, {\it Dynamics in nonlocal cosmological models derived from string field theory}
  \prd{76}{105007}{2007};

   L. Joukovskaya, {\it Rolling tachyon in nonlocal cosmology} AIP
Conf.Proc. 957 325-328,2007,  [\hepth{0710.0404}];


 L.V. Joukovskaya, {\it  Dynamics with Infinitely Many Time Derivatives in Friedmann-Robertson-Walker Background and Rolling Tachyon},
 \jhep{02}{2009}{045}, [\hepth{0807.2065}].


\bibitem{AK-dil} I.Ya. Aref'eva , A.S. Koshelev,
 {\it Cosmological Signature of Tachyon
Condensation},  \jhep{0809}{068}{2008}, [\hepth{0804.3570}]
 \bibitem{BMNR} N. Barnaby, D. J. Mulryne, N. J. Nunes, P.
Robinson, {\it Dynamics and Stability of Light-Like Tachyon Condensation},
[\hepth{0811.0608}]
 \bibitem{Hellerman-MS}
S.~Hellerman and M.~Schnabl, {\it Light-like tachyon condensation in open
  string field theory}, [\hepth{0803.1184}].

\end{thebibliography}
\end{document}